\documentclass[]{aa}
\usepackage[varg]{txfonts}

\usepackage{natbib}
\bibpunct{(}{)}{;}{a}{}{,} 
\usepackage{graphicx}
\usepackage{color}
\usepackage{amstext}
\usepackage{subfigure}
\newcommand{\angstrom}{\mbox{\normalfont\AA}}

\begin{document}

\title{Nitrogen abundance in the X-ray halos of clusters and groups of galaxies}

\author{Junjie Mao\inst{\ref{inst1},~\ref{inst2}}, 
        J. de Plaa\inst{\ref{inst1}}, 
        J. S. Kaastra\inst{\ref{inst1},~\ref{inst2}},
        Ciro Pinto\inst{\ref{inst3}},
        Liyi Gu\inst{\ref{inst1}},
        F. Mernier\inst{\ref{inst1},~\ref{inst2}}, 
        Hong-Liang Yan\inst{\ref{inst4}},
        Yu-Ying Zhang\inst{\ref{inst5}}, 
        \and 
        H. Akamatsu\inst{\ref{inst1}}
        }

\offprints{J. Mao,~\email{J.Mao@Sron.nl}}

\institute{SRON Netherlands Institute for Space Research, Sorbonnelaan 2, 3584 CA Utrecht, the Netherlands \label{inst1}
\and Leiden Observatory, Leiden University, Niels Bohrweg 2, 2300 RA Leiden, the Netherlands \label{inst2}
\and Institute of Astronomy, Madingley Road, CB3 0hA Cambridge, UK \label{inst3}
\and Key Laboratory of Optical Astronomy, National Astronomical Observatories, Chinese Academy of Sciences, 20A Datun Road, Chaoyang District, 100012 Beijing, PR China \label{inst4}
\and Argelander-Institut f\"{u}r Astronomie, Universit\"{a}t Bonn, Auf dem H\"{u}gel 71, 53121 Bonn, Germany \label{inst5}
}

\date{Received date / Accepted date}

\abstract
{Chemical abundances in the X-ray halos (also known as the intracluster medium, ICM) of clusters and groups of galaxies can be measured via prominent emission line features in their X-ray spectra. Elemental abundances are footprints of time-integrated yields of various stellar populations that have left their specific abundance patterns prior to and during the cluster and group evolution.}
{We aim to constrain nitrogen abundances in the CHEmical Evolution RGS Sample (CHEERS), which contains 44 nearby groups and clusters of galaxies,
in order to have a better understanding of their chemical enrichment.}
{We examine the high-resolution spectra of the CHEERS sample carefully and take into account various systematic effects in the spectral modelling. We compare the observed abundance ratios with those in the Galactic stellar populations, as well as predictions from stellar yields (low- and intermediate-mass stars,
massive stars and degenerate stars). }
{The nitrogen abundance can only be well constrained ($\gtrsim3\sigma$) in one cluster of galaxies and seven groups of galaxies. The [O/Fe] -- [Fe/H] relation of the ICM is comparable to that for the Galaxy, while both [N/Fe] and [N/O] ratios of the ICM are higher than in the Galaxy. Future studies on nitrogen radial distributions are required to tell whether the obtained higher [N/Fe] and [N/O] ratios are biased due to the small extraction region ($r/r_{500}\lesssim0.05$) that we adopt here. Since abundances of odd-$Z$ elements are more sensitive to the initial metallicity of stellar populations, accurate abundance measurements of N, Na and Al are required to better constrain the chemical enrichment in the X-ray halos of clusters and groups of galaxies.}
{}

\titlerunning{Nitrogen abundance in the chemical evolution RGS sample}
\authorrunning{Mao et al.}
\maketitle

\section{Introduction}
\label{sct:intro}
Clusters of galaxies aggregate baryons and dark matter within large-scale structures that have collapsed under their own gravity. A large fraction ($\sim$15$-$20\%) of the total mass of a cluster is in the hot (T$\sim$$10^{7-8}$~K) X-ray halos (also known as the intracluster medium, ICM), while the member galaxies only make up for $\sim$3$-$5\% of the total mass. The rest is in the form of dark matter. The ICM is an attractive laboratory for the study of nucleosynthesis and chemical enrichment \citep[for a review, see][]{wer08}. Due to its deep gravitational potential well, a massive cluster \citep[$M\gtrsim 10^{13}~{\rm M_\odot}$,][]{ren14} can be considered as a ``closed-box" \citep[e.g.][]{whi93}, i.e. all the metals synthesized by different stellar populations in the member galaxies are conserved within the cluster. This assumption is based on the consistency  \citep{lan13} between the total cluster baryon fraction within a certain radius, say $r_{500}$\footnote{The radius within which the plasma mass density is 500 times the critical density of the Universe at the redshift of the groups and clusters of galaxies.}, and the cosmic baryon fraction. This assumption does not necessarily hold for less massive groups of galaxies, due to the relatively shallow gravitational potential well. Once the metals are released via stellar winds, supernovae, etc., various metal transportation mechanisms working on different locations and time-scales distribute the metals into the X-ray halos. Relevant metal transportation mechanisms include Galactic winds, ram--pressure stripping, AGN--ICM interaction, and galaxy--galaxy interaction \citep[for a review, see][]{sch08}.

Assuming that stellar populations where the metals are synthesized are representative, given stellar yields and observed abundance patterns in the X-ray halos,
we are able to put constrains on the chemical enrichment to the largest gravitationally bound systems in the Universe \citep[for a review, see][]{boh10}. Given that $\alpha$ elements (e.g. O, Ne, Mg) are mainly produced in massive stars via core-collapse supernovae (SNcc), and Fe-peak elements are mainly produced 
in degenerate stars via Type Ia supernovae (SNIa), the SNIa fraction with respect to the total number of supernovae (SNcc plus SNIa) 
that enriched the ICM can be obtained by either fitting \citep[e.g.][]{wer06b, dpl06, san06, sat07, kom09} the best-fit elemental abundances with supernova yields or applying directly supernova yields to immediately predict the X-ray spectrum \citep{bul12}. The latter assumes that the ICM can be described as a single temperature plasma in collisional ionized equilibrium (CIE). Unlike elements heavier than oxygen, carbon and nitrogen are mainly produced in low- and intermediate-mass stars \citep[for a review, see][]{nom13}. Thus, the ICM abundances of C and N also provide important information to better understand the chemical enrichment.

Observationally, the abundances of $\alpha$ elements and Fe-peak elements can be measured with both low- and high-resolution grating spectra \citep[e.g.,][]{mer16a, dpl17,hit17a}. Carbon and nitrogen abundances can only be determined from high-resolution grating spectra, such as those obtained with XMM--{\it Newton}/RGS \citep[Reflection Grating Spectrometers,][]{dhe01}. \citet{xu02} first reported the nitrogen abundance in the hot X-ray halo of NGC\,4636. Later, the nitrogen abundance was reported in other individual targets \citep{tam03, wer06a, san08, wer09, san10, gra11} and in the stacked spectra of 62 groups and clusters of galaxies in \citet{san11a}. 

In this work, we systematically study the nitrogen abundance in the CHEERS sample\footnote{CHEERS is short for CHEmical Evolution Rgs cluster Sample.}\citep{dpl17}, which contains 44 nearby ($z<0.1$) X-ray bright cool-core groups and clusters of galaxies. The key sample selection criterion \citep{dpl17} of the CHEERS sample is that the \ion{O}{viii} Ly$\alpha$ line at $\sim 18.97~\angstrom$ (rest frame) is detectable ($\gtrsim5\sigma$) with RGS. 

Throughout the paper we use $H_0 = 70~{\rm km~s^{-1}~Mpc^{-1}}$, $\Omega_{\rm M} = 0.3$, $\Omega_{\Lambda} = 0.7$. For the spectral analysis (Section~\ref{sct:spec_mo}), we use $C$-statistics following \citet{kaa17a}. Unless specified otherwise, all errors correspond to the 68\% confidence level for one interesting parameter.

\section{Data reduction}
\label{sct:dr}
We reduced both RGS and EPIC/MOS data following the same procedures described in \citet{pin15}, using XMM-{\it Newton} Science Analysis System\footnote{http://www.cosmos.esa.int/web/\textit{XMM-Newton}/sas} (SAS) v15.0.0. MOS data are reduced since the Reflection Grating Assemblies (RGAs) are aligned with the light path of the MOS cameras. We use MOS data for screening soft-proton flares and deriving the spatial extent of the source along the dispersion direction of RGS. 

For each observation, we extract RGS spectra in a $\sim$3.4-arcmin-wide (along the cross-dispersion direction) region centred on the emission peak. This is done by setting the {\it xpsfincl} mask to include 99\% of the line spread function (LSF) inside the spatial source extraction mask. The extraction region is somewhat different from the circular aperture used for the EPIC data analysis, especially when there is a gradient in temperature structure and/or metal abundances. The spectra and response matrices are converted to SPEX \citep{kaa96} format through the SPEX task {\it trafo}. The RGS modelled background spectra are subtracted.

The spatial extent along the dispersion direction of the source dominates the broadening of the emission lines, which can be described as \citep{tam04}
\begin{equation}
\Delta \lambda = \frac{0.138}{m} \frac{\Delta \theta}{\rm arcmin} \angstrom,
\label{eq: dlam}
\end{equation}
where $m$ is the spectral order, $\Delta \theta$ is the offset angle of the source. The average spatial extent of the ICM that includes half of the maximum line flux 
is $\sim$2', that is to say, the average FWHM of the line profile (Equation~\ref{eq: dlam}) is $\sim$0.276~$\angstrom$ (1st-order) and $\sim$0.138~$\angstrom$ (2nd-order), respectively. The bin size that we used in our data processing with {\it rgsproc} is 0.01~$\angstrom$ (1st-order) and 0.005~$\angstrom$ (2nd-order), respectively. Hence we re-binned the RGS spectra by a factor of 10 for both 1st-order (7$-$28~$\angstrom$) and 2nd-order data (7$-$14~$\angstrom$), which approximately yielded the optimal binning \citep[1/2$-$1/3 FWHM, ][]{kaa16} for RGS spectra of the ICM. 

\section{Spectral analysis}
\label{sct:spec_mo}
The high-resolution X-ray spectral analysis package SPEX (v3.03) is used to fit the RGS spectra. For collisional ionized equilibrium (CIE) plasma modelling, a large portion of the out-dated atomic data from the old version of SPEX (v.2.05) has been replaced with the state-of-the-art results published in the last decade, such as level-resolved radiative recombination data \citep{badn06, mao16} and ionization balance that includes inner-shell ionization data \citep{urd17}. In addition, atomic data including collisional excitation/de-excitation rates, radiative transition probabilities and auto-ionization rates have been consistently calculated using the FAC\footnote{https://www-amdis.iaea.org/FAC} code \citep{gu08} and are included in the latest version of SPEX code as well. The \textit{Hitomi} Soft X-ray Spectrometer (SXS) spectrum of the Perseus cluster offers an unprecedented benchmark of popular atomic codes, we refer to \citet{hit17b} for more details.

For each cluster or group of galaxies, we fit simultaneously RGS1 and RGS2 spectra of each observation. Unless specified otherwise, the redshifts and Galactic absorption column densities are frozen to the values given in \citet{pin15}. We use the collisional ionization equilibrium absorption model \citep{dpl04, ste05} with a fixed temperature $T=0.5$~eV to account for the Galactic neutral absorption. When modelling the thermal component(s) of the ICM, we consider three different differential emission measure (DEM) distributions.
\begin{enumerate}
\item The simplest scenario is assuming that the ICM is isothermal so that it can be described as a single temperature CIE model (denoted as 1T).
\item A more complicated scenario is that the ICM consists of a hotter and a cooler CIE component (denoted as 2T). Abundances of the two thermal components are assumed to be the same, while emission measures and temperatures are free to vary.
\item The most sophisticated scenario requires a multi-temperature DEM distribution. We adopted the GDEM model \citep{dpl06} here, which assumed a Gaussian distribution of the DEM in $\log~T$,
\begin{equation}
Y(x)= \frac{Y_0}{\sigma \sqrt{2\pi}} {\exp}\left(-\frac{(x-x_0)^2}{2\sigma^2}\right)~,
\label{eq: gdem}
\end{equation}
where $x =\log(T)$ and $x_0=\log(T_0)$, with $T$ and $T_0$ (peak temperature of the distribution) in units of keV,  and $Y_0$ is the emission measure. Apparently, when $\sigma=0$, GDEM is identical to 1T. Again, abundances of the multi-temperature components are assumed to be the same.
\end{enumerate}
The above three DEM distributions are driven by the results of \citet{fra13}, where the authors measured ${\rm DEM}=dY/dT$ (where $Y=\int n_{\rm e} n_{\rm H} dV$ is the emission measure) distribution of 62 galaxy clusters in the HIFUGCS sample \citep{zha11}. By comparing the goodness of the fit, one of the DEM distributions is favored and reported for each cluster or group. Regardless of the choice of the DEM distribution, the abundances of N, O, Ne, Mg, Fe, and Ni are free to vary, while the other elements heavier than He are frozen to 0.3~solar \citep[e.g.][]{fuj08, wer13}. All the abundances are normalized to the proto-solar abundances of \citet{lod09}, i.e. $z_{i,~\rm ICM}/z_{i,~\rm \odot}$. The ionization balance described in \citet{urd17} is used. The spatial broadening is taken into account by convolving the thermal plasma model (1T/2T/GDEM) with the spatial broadening model ({\it lpro}).

The general strategy mentioned above does not necessarily provide an accurate measurement of elemental abundances. Special treatments are required in some cases. When $N_{\rm H~I}\gtrsim 7\times10^{24}~{\rm m^{-2}}$, the Galactic hydrogen column density (denoted as NH)  is allowed to vary \citep{dpl17}. When the ICM thermal emission is contaminated by non-thermal emission, a power law component is added accordingly with parameters fixed to literature values \citep[e.g. for M\,87 see][]{wer06a}. The derived abundance of a given element is proportional to the equivalent width, i.e. the ratio between the line flux and the continuum flux, given that the abundance is determined mainly from a well resolved emission line. That is to say, any uncertainty in the continuum would also impact the abundance measurement. When fitting the RGS spectra for a broad wavelength range (7--28~$\angstrom$ in our case), the continuum flux may be slightly over- or under-estimated due to uncertainties in the calibration of the RGS effective area, background subtraction, etc. Consequently, abundance measurement might be significantly biased, compared to the statistical uncertainties in the spectral fit. The top-left panel of Figure~\ref{fig:cheers_spec_plot_zoom} shows that the global fit overestimates the nitrogen abundance in M\,87. A similar issue was pointed out by \citet{mer15} for their EPIC spectral analysis and the authors performed a local fit around a specific line of interest to improve the accuracy of the abundance measurement. Here we also performed the local fit ($\pm1~\angstrom$ around the line centre) to check whether the global continuum level is correct or not, if not, the local fit results are adopted. For instance, while the global fit overestimates the N/Fe ratio ($2.9\pm0.3$) for M\,87, the local fit yields a more accurate N/Fe ratio ($1.8\pm0.2$). Other systematic uncertainties regarding the spatial broadening of the line (Appendix~\ref{sct:lpro}) and RGS background model (Appendix~\ref{sct:mbkg}) can be found in the Appendix \ref{sct:sys_err}.

\section{Results and comparison with literature values}
\label{sct:spec_res}
\begin{figure*}
\centering
\includegraphics[width=\hsize, trim={1cm 2cm 1cm 2cm}, clip]{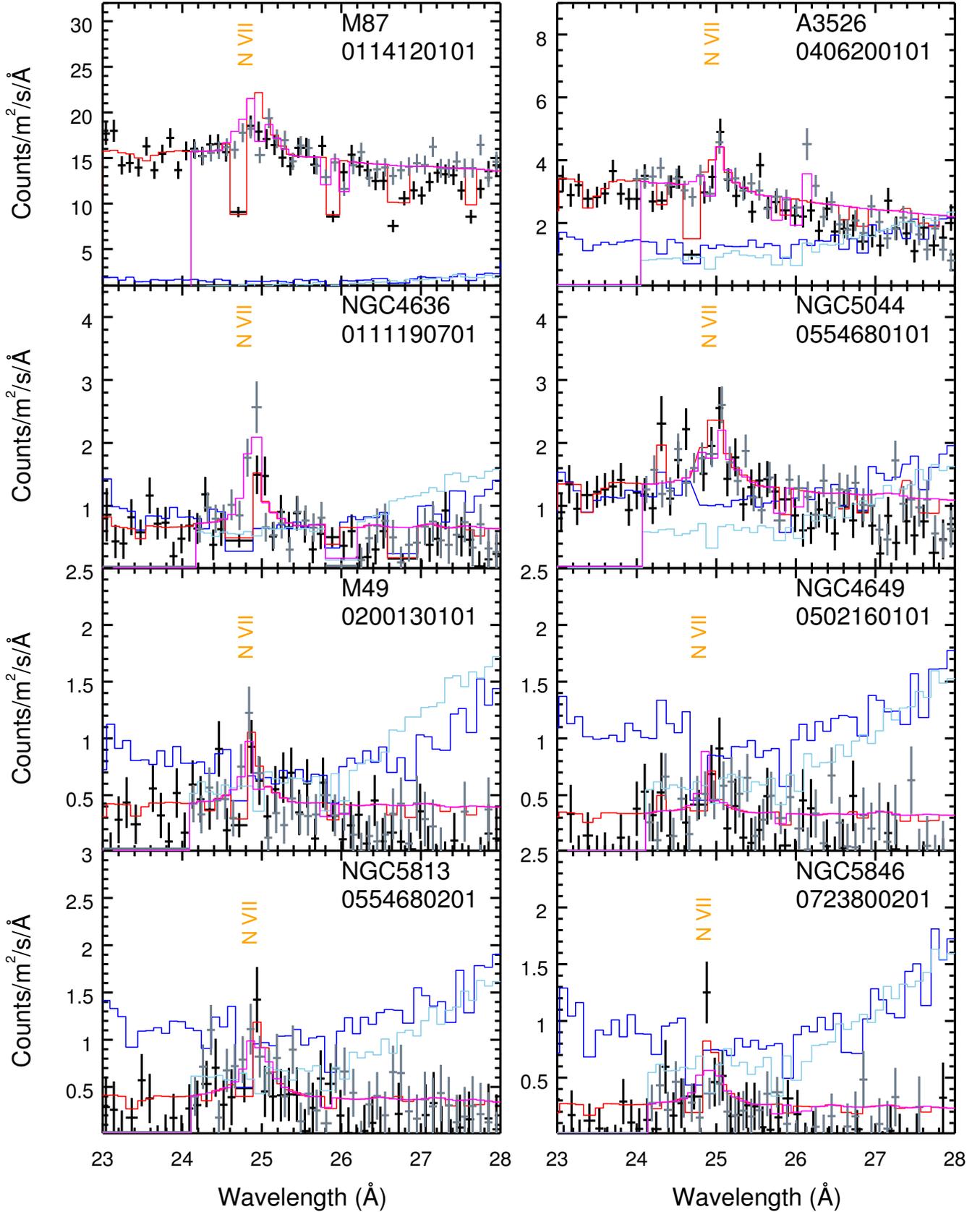}
\caption{Observed RGS spectra in the \ion{N}{VII} neighbourhood, i.e., 23--28 $\angstrom$ wavelength range. The data points are shown in black (RGS1) and grey (RGS2), the modelled background spectra are shown in deep blue (RGS1) and light blue (RGS2) histograms, and the (global fit) model spectra are shown in red (RGS1) and magenta (RGS2) histograms. Vertical dashed lines indicate the \ion{O}{VIII} Ly$\alpha$ around 19~\AA\ and \ion{N}{VII} Ly$\alpha$ around 25~\AA\ in the observed frame. Spectra from merely one observation per target are shown for clarity. Note that in the global fit to M\,87 (the top-left panel), the N abundance is clearly overestimated, thus a local fit is performed (Section~\ref{sct:spec_mo}) to obtain a more accurate abundance measurement.} 
\label{fig:cheers_spec_plot_zoom}
\end{figure*}

Nitrogen abundance measurements are best done in plasma with lower temperature (Figure~\ref{fig:icon_occ}). Therefore, in the CHEERS sample, we found that the nitrogen abundance can merely be well constrained ($\gtrsim3\sigma$) in the core ($r/r_{500} \lesssim0.05$) of one cluster of galaxies (\object{A\,3526}) and seven groups of galaxies (\object{M\,49}, \object{M\,87}, \object{NGC\,4636}, \object{NGC\,4649}, \object{NGC\,5044}, \object{NGC\,5813}, \object{NGC\,5846}). For some of the lower temperature sources (e.g., NGC\,3411) in the CHEERS sample, more exposure time is required to better constrain the nitrogen abundance.  Spectral fits near the \ion{N}{vii} Ly$\alpha$ line for these eight targets are shown in Figure~\ref{fig:cheers_spec_plot_zoom} and the same (global) fits to the $7-28$ \AA\ wavelength range can be found in Figure~\ref{fig:cheers_spec_plot}. The abundances and abundance ratios are summarized in Table~\ref{tbl:icm_all}.

\begin{table*}
\caption{Abundances and abundance ratios within the 3.4-arcmin-wide (along cross dispersion direction) extraction region. }
\label{tbl:icm_all}
\centering
\begin{tabular}{cccccccccccccccccccccc}
\hline\hline
Source & A3526 & M49 & M87 & NGC4636 & NGC4649 & NGC5044 & NGC5813 & NGC5846 \\  
\hline\hline
\noalign{\smallskip}
$r/r_{500}$ & 0.026 & 0.018 & 0.012 & 0.022 & 0.015 & 0.034 & 0.031 & 0.036 \\
\noalign{\smallskip}
kpc & 43.2 & 18.7 & 17.7 & 15.6 & 15.6 & 37.7 & 26.9 & 25.8 \\
\noalign{\smallskip}
Model & NH+2T & GDEM & 2T+PL & 2T & 1T & GDEM & 2T & 2T \\
\noalign{\smallskip}
$C$-stat./d.o.f. & 2186/1088 & 852/544 & 3954/1111 & 748/480 & 1530/1096 & 1670/1094 & 2480/1649 & 1825/1093 \\
\noalign{\smallskip}
$\sigma_{\rm N/Fe}$ & $\sim7\sigma$ & $\sim3\sigma$ & $\sim9\sigma$ & $\sim4\sigma$ &  $\sim3\sigma$ & $\sim5\sigma$ & $\sim5\sigma$ & $\sim3\sigma$ \\ 
N/O & $2.7\pm0.5$ & $2.7\pm1.0$ & $2.2\pm0.3$ & $3.3\pm1.1$ & $2.9\pm1.0$ & $2.2\pm0.5$ & $3.2\pm0.9$ & $2.7\pm0.8$ \\
\noalign{\smallskip}
N/Fe & $1.5\pm0.2$ & $1.6\pm0.6$ & $1.8\pm0.2$ & $1.9\pm0.5$ & $2.4\pm0.8$ & $1.4\pm0.3$ & $1.9\pm0.4$ & $2.3\pm0.7$ \\
\noalign{\smallskip}
O/Fe & $0.54\pm0.04$ & $0.59\pm0.10$ & $0.82\pm0.03$ & $0.59\pm0.08$ & $0.84\pm0.11$ & $0.65\pm0.05$ & $0.58\pm0.07$ & $0.86\pm0.12$ \\
\noalign{\smallskip}
Ne/Fe & $0.57\pm0.06$ & $0.66\pm0.17$ & $0.55\pm0.05$ & $0.64\pm0.12$ & $1.07\pm0.19$ & $0.68\pm0.08$ & $0.53\pm0.09$ & $0.71\pm0.14$ \\
\noalign{\smallskip}
Mg/Fe & $0.66\pm0.07$ & $0.79\pm0.19$ & $0.24\pm0.04$ & $0.64\pm0.13$ & $1.40\pm0.23$ & $0.77\pm0.08$ & $0.83\pm0.11$ & $0.66\pm0.14$ \\
\noalign{\smallskip}
Fe & $1.02\pm0.03$ & $1.50\pm0.12$ & $0.55\pm0.01$ & $0.66\pm0.04$ & $0.55\pm0.03$ & $0.78\pm0.03$ & $0.92\pm0.04$ & $0.77\pm0.05$ \\
\noalign{\smallskip}
Ni/Fe & $1.2\pm0.1$ & $1.8\pm0.5$ & $0.65\pm0.07$ & $2.0\pm0.4$ & $2.5\pm0.4$ & $1.5\pm0.3$ & -- -- & $2.0\pm0.4$ \\
\noalign{\smallskip}
\hline
\end{tabular}
\tablefoot{Abundances and abundance ratios are given according to the proto-solar abundance of \citet{lod09}. Statistical uncertainties ($1\sigma$) are quoted here. Systematic uncertainties on the abundance ratios are estimated in Section~\ref{sct:spec_res}. $\sigma_{\rm N/Fe}$ is the significance level of nitrogen detection according to the N/Fe ratio (to be greater than zero). The uncertainties shown are 1$\sigma$ statistical error bars. 1T, 2T and GDEM refer to single-temperature, two-temperature and multi-temperature differential emission measure (DEM) distribution (Section~\ref{sct:spec_mo}). For A\,3526, ``NH" refers to a free Galactic hydrogen column density in the spectral analysis. The Galactic hydrogen column densities for the other seven systems are frozen to literature values. For M\,87, we use a power-law (PL) to model the non-thermal component, which is variable between the two observations \citep{wer06a}. For NGC\,5813, Ni abundance cannot be constrained, and we fix it to solar during the fitting.}
\end{table*}

We notice that M\,87 and A\,3526 have the worst $C$-stat. over degrees of freedom (d.o.f.) ratio (Table~\ref{tbl:icm_all}), but the poor statistics do not have a significant impact on our astrophysical interpretation of the measured abundance ratios. For instance, there are some ``features" around $27-28$~\AA\ for M\,87, and the two RGS instruments do not agree with each other. Since the source is bright and has a deep exposure, the aforementioned ``features" are all statistically significant thus contributing to the poor $C$-stat. over d.o.f. ratio. Such ``features" are mainly caused by the imperfect instrument calibration on the effective area. Nevertheless, as shown in Figure~\ref{fig:cheers_spec_plot}, the global fit yields a good estimate of the continuum and the astrophysical features of the spectrum. Furthermore, we use the local fit to correct biased abundance measurements in the global fit.

The O/Fe ratios in the above eight sources (with the extraction region of $\sim$3.4 arcmin) are $\lesssim1.3$. Some of our results differ from those reported in \citet{dpl17} for the 0.8-arcmin-wide extraction region. This is mainly due to the strong temperature and abundance gradients \citep{mer17a}, if present.

The N/Fe ratios reported in Table~\ref{tbl:icm_all} are $\gtrsim1.4$ with larger scatter. \citet{xu02},  \citet{wer06a, wer09} and \citet{gra11} also reported similar N/Fe ratio $\gtrsim 1.4$ with large scatter between individual targets. \citet{tam03} reported a lower N/Fe ratio of $\sim(0.6\pm0.2)$ for \object{NGC\,5044}. The N/Fe ratios for individual targets reported in \citet{san08, san10} are (spectral fitting) model dependent, with both higher ($\gtrsim$1) and lower ($\lesssim$1) values. 

The N/O ratios reported here are above zero at the $\gtrsim2.5\sigma$ confidence level. \citet{san11a} reported a N/O ratio of $4.0\pm0.6$ in the stacked spectra of 62 groups and clusters of galaxies. Note that the authors also point out that, among individual targets in their sample, the nitrogen abundances vary considerably. 

We caution that the reported abundance ratios in the literature depend on the details of the spectral analysis and/or the adopted extraction region. As discussed in great detail \citep{dpl17} the O/Fe ratio can be biased up to $30\%$ (in total) due to various kinds of systematic uncertainties, including the effect of spatial line broadening, the choice of the multi-temperature model, the influence of the assumed value of galactic hydrogen column density $(N_{\rm H})$. In addition, the background level around the \ion{N}{vii} Ly$\alpha$ neighbourhood is rather high in many cases (Figure~\ref{fig:cheers_spec_plot_zoom}). But the N/Fe ratio is not affected by the uncertainties (a $\lesssim10\%$ constant bias) in the RGS modelled background (Appendix~\ref{sct:mbkg}). As we mentioned in Section~\ref{sct:spec_res}, we perform local fit in some cases (e.g. M\,87) to mitigate some systematic uncertainties. In short, the overall systematic uncertainties of the N/Fe ratio is expected to be within 30\%. On the other hand, the overall systematic uncertainties of the Ne/Fe, Mg/Fe and Ni/Fe ratios are expected to be larger than 30\%. A further investigation is required to carefully estimate the systematic uncertainty of these abundance ratios, which is beyond the scope of this paper.

In addition, \citet{smi09} reported ${\rm N/Fe}\lesssim3$ after analyzing optical spectra of $\sim$150 red-sequence galaxies over a wide mass range in the Coma cluster and the Shapley Supercluster. \citet{joh12} found ${\rm N/Fe}\lesssim3$ for $\sim$4000 early-type galaxies in a narrow redshift bin $z\in(0.05,~0.06)$ observed with Sloan Digital Sky Survey (SDSS). In contrast, \citet{gre15} measured a remarkably super-solar (at least three times solar) N/Fe abundance within a radius of 15 kpc for $\sim$100 massive early-type galaxies. As pointed out by \citet{gre15}, their N/Fe ratio would be three times solar if the O/Fe ratio were solar. In their default analysis, Greene et al. assume [O/Fe]$=$0.5, which leads to even higher N/Fe ratio. This assumed value of [O/Fe] is higher (by 0.2 dex or so) than some chemical enrichment model predicted \citep{pip09, nom13}. A full spectral modelling is required to mitigate the impacts of blending and the uncertainties introduced by oxygen \citep{gre15}. Anyway, the O/Fe ratio assumed/predicted in the optical analysis is higher than that observed in the X-ray wavelength range. But it is possible that the SNcc products are preferably locked up by stars \citep{loe13}. In short, it is not trivial to compare and interpret the abundance ratios measured in the X-ray wavelength range and the optical wavelength range.

Moreover, The Ni/Fe abundance ratios reported in Table~\ref{tbl:icm_all} differ from the solar Ni/Fe ratio found in the Perseus cluster \citep{hit17a}. This might be due to the fact that the present work uses the L-shell lines which have large uncertainties in the current atomic codes.

\section{Discussion}
\label{sct:dis}
In the Galactic chemical evolution model \citep[e.g.][]{kob06, nom13}, nitrogen is mainly enriched via stellar winds of low- and intermediate-mass stars in the asymptotic giant branch (AGB). Therefore, in this section, we include the AGB enrichment channel (Section~\ref{sct:icm_ce}) for the chemical enrichment theory \citep[e.g.][]{loe13}. We then compare the [O/Fe] -- [Fe/H]\footnote{$\left[{\rm A}/{\rm B} \right]_{\rm ICM / star} = \log_{10}\left(\frac{N_{\rm A}}{N_{\rm B}}\right)_{\rm ICM / star} - \log_{10}\left(\frac{N_{\rm A}}{N_{\rm B}}\right)_{\odot} =\log_{10}\left(\frac{Z_{\rm A}}{Z_{\rm B}}\right)_{\rm ICM / star}$.} and [N/Fe] -- [Fe/H] relation between the ICM and different Galactic stellar populations,  as well as the [N/O] -- [O/Fe] relation between the ICM and supernova yields (Section~\ref{sct:origin}). These comparisons enable us to discuss whether the nitrogen enrichment in the ICM shares the same origin as that in the Galaxy. Finally, we study elemental abundances in NGC\,5044 (Section~\ref{sct:odd_z}) to illustrate that by including odd-$Z$ elements like nitrogen, the initial metallicity of the stellar population that enriched the ICM can be better constrained.

\subsection{ICM Chemical enrichment}
\label{sct:icm_ce}
To interpret the observed time-integrated chemical abundances, we assume a single population of stars formed at high redshift \citep[say, $z=2-3$,][]{hen15} with a common initial mass function (IMF). The ICM elemental abundance (the number of atoms of the $i$th element relative to that of hydrogen) relative to solar is defined as 
\begin{equation}
Z_{i,~\rm ICM}=\frac{z_{i,~\rm ICM}}{z_{i,~\odot}}=\frac{N_{\rm ICM}^{\rm li} <y_i^{\rm li}> + N_{\rm ICM}^{\rm m} <y_i^{\rm m}>+N_{\rm ICM}^{\rm d}y_i^{\rm d}}{M_{\rm ICM} X (A_i/A_H) (n_{i,~\odot}/n_{\rm H,~\odot})}~,
\label{eq:z_icm}
\end{equation}
where $N_{\rm ICM}^{li/m/d}$ are the total number of low- and intermediate-mass stars (denoted with superscript ``li") that enrich the ICM via the AGB channel, massive stars (``m") that explode as SNcc or PISNe (pair-instability supernovae), and single/double degenerate (``d") stars that explode as SNIa to enrich the ICM, $y^{\rm li/m/d}$ the corresponding stellar yields, ${M_{\rm ICM}}$ the mass of the ICM, $A_i$ the atomic weight of the $i$th element, $A_{\rm H}=1.0086~a.m.u.$ the atomic weight of H, $n_{i, \odot}$ the elemental abundance by number in the solar abundance table and $X$ is the mass fraction of H in the present universe. 

The first two terms in the numerator of Equation~(\ref{eq:z_icm}) include the IMF weighted yields of low- and intermediate-mass or massive stars
\begin{equation}
<y_i>=\frac{\int_{m_{\rm lo}}^{m_{\rm up}} \phi(m) y_i(m) dm}{\int_{m_{\rm lo}}^{m_{\rm up}} \phi(m) dm}~,
\label{eq:y_lim}
\end{equation}
where $\phi(m)$ is the IMF,  $m_{\rm lo}$ and $m_{\rm lo}$ the lower and upper mass limit ($Z_{\rm init}$ dependent, Table~\ref{tbl:mr_lim}) of low- and intermediate-mass or massive stars considered.

\begin{table}
\caption{The mass ranges (in $M_{\odot}$) used for calculating the IMF-weighted yields (Equation~\ref{eq:y_lim}). For both low/intermediate-mass stars (``li", progenitors of AGBs) and massive stars (``m", progenitors of SNcc and PISNe), the mass ranges depend on the initial metallicity ($Z_{\rm init}$) of the stellar population.}
\label{tbl:mr_lim}
\centering
\begin{tabular}{cccccccccccccccccccccc}
\hline\hline
$Z_{\rm init}$ & $(m_{\rm lo}, ~m_{\rm up})^{\rm li}$ & $(m_{\rm lo}, ~m_{\rm up})^{\rm m}$ \\
\hline\hline
0 & (0.9,~3.5) & (11,~300)  \\
0.001 & (0.9,~6.5) & (13,~40)  \\
0.004 & (0.9,~6.5) & (13,~40)  \\
0.008 & (0.9,~6.5) & (13,~40)  \\
0.02 & (0.9,~7.0) & (13,~40)  \\
0.05 & (0.9,~7.0) & (13,~40)  \\
\hline
\end{tabular}
\tablefoot{The upper mass limit of intermediate-mass stars, defined as the minimum mass for the off-centre carbon ignition to occur, is smaller for lower metallicity \citep{ume02,gil07,sie07}. The upper mass limit of massive stars depends on the types of supernovae that are taken into account. Massive stars that explode as core-collapse supernovae (with $m_{\rm up}=40~M_{\odot}$ for $Z\ne 0$ and an explosion energy of $10^{44}~{\rm J}$) and pair-instability supernovae (with $m_{\rm up}=300~M_{\odot}$ and an explosion energy greater than $10^{44}~{\rm J}$) are considered here. }
\end{table}

It is unclear whether a universal IMF is applicable to all the clusters and groups of galaxies, or that the IMF depends on the local star formation rate (SFR) and/or 
metallicity of the environment \citep{mol15}. For simplicity, other than the standard Salpeter IMF, we consider a top-heavy IMF, which is probably more relevant here, with an arbitrary IMF index (unity here). We caution that changing the IMF has profound consequences \citep{rom05, pol12}, including observables other than the chemical abundances that we measured here. For instance, a top-heavy IMF might make the galaxies too red \citep{sar06}. The global impacts on other observables introduced by the non-standard IMF are beyond the scope of this paper.

Additionally, the IMF weighted yield for massive stars, $y_i^{\rm m}$, depends on the type(s) of supernovae that are taken into account for massive stars.
We consider massive stars with stellar mass between 10~$M_{\odot}$ and 40~$M_{\odot}$ ($Z_{\rm init}>0$) or 140~$M_{\odot}$ ($Z_{\rm init} = 0$)
that undergo Fe core collapse at the end of their evolution and become Type II and Ib/c supernovae (i.e. core-collapse supernovae). Massive stars in the mass range of 25~$M_{\odot}$ to 40~$M_{\odot}$ ($Z_{\rm init}>0$) or 140~$M_{\odot}$ ($Z_{\rm init} = 0$) can alternatively give rise to hypernovae (HNe) or faint supernovae (FSNe), instead of normal SNcc. Since the ratios among normal SNcc, HNe and FSNe for the relevant  mass range are unknown for clusters and groups of galaxies, we do not consider HNe and FSNe enrichment for simplicity. In addition, we also take into account pair-instability supernovae \citep{ume02} 
for zero initial metallicity ($Z_{\rm init}=0$) enrichment, assuming that all the very massive stars, with stellar mass between 140~$M_{\odot}$ and 300~$M_{\odot}$, undergo pair-instability supernovae\footnote{If very massive stars do not lose much mass, they are completely disrupted without forming a black hole via pair-instability supernovae \citep{bar67}.} (PISNe). Therefore, our calculation of the predicted abundance (Equation~\ref{eq:z_icm}) is a first-order approximation. 

The last term in the numerator of Equation~(\ref{eq:z_icm}) include $y_i^{\rm d}$, which is the yield per SNIa, and depends on the SNIa model. SNIa yields from \citet{iwa99}, \citet{bade06} and \citet{mae10} are used for the following analysis. 

In Table~\ref{tbl:lim_idx}, we summarize the 12 sets of IMF weighted yields for non-degenerate stars that enrich the ICM via AGBs, SNcc (and PISNe). In Table~\ref{tbl:snia_idx} we summarize the 16 sets of SNIa yields for degenerate stars that enrich the ICM via SNIa.

\begin{table}
\caption{Summary of the underlying model dependency for IMF power-law index and initial metallicity ($Z_{\rm init}$) of the stellar population (Equation~\ref{eq:y_lim}).}
\label{tbl:lim_idx}
\centering
\begin{tabular}{cccccccccccccccccccccc}
\hline\hline
Index & (IMF, $Z_{\rm init}$) & Index & (IMF, $Z_{\rm init}$)  \\
\hline\hline
1 & (2.35,~0.0) & 7 & (1.0,~0.0) \\
2 & (2.35,~0.001) & 8 & (1.0,~0.001) \\
3 & (2.35,~0.004) & 9 & (1.0,~0.004) \\
4 & (2.35,~0.008) & 10 & (1.0,~0.008) \\
5 & (2.35,~0.02) & 11 & (1.0,~0.02) \\
6 & (2.35,~0.05) & 12 & (1.0,~0.05)\\
\hline
\end{tabular}
\end{table}
\begin{table}
\caption{Summary of the underlying model dependency for SNIa enrichment (Equation~\ref{eq:z_icm}).}
\label{tbl:snia_idx}
\centering
\begin{tabular}{cccccccccccccccccccccc}
\hline\hline
Index & Model & Index & Model \\
\hline\hline
1 & CDD1 & 2 & CDD2 \\
3 & W7 & 4 & W70 \\
5 & WDD1 & 6 & WDD2 \\
7 & WDD3 & 8 & DDTa  \\
9 & DDTb & 10 & DDTc \\
11 & DDTd & 12 & DDTe \\
13 & DDTf & 14 & CDEF \\
15 & ODDT & 16 & CDDT \\
\hline
\end{tabular}
\tablefoot{$^{a}$ The IMF power-law index and the initial metallicity ($Z_{\rm init}$) of the stellar population. $^{b}$ SNIa models. The CDD (i.e. index 1 and 2) and WDD (5 to 7) models are delayed-detonation scenario \citep{iwa99}. The W (3 and 4) models refer to convection deflagration scenario \citep{iwa99}. The DDT (8 to 12) models are based on observational results from the Tycho supernova remnant \citep{bade06}. The CDEF model refers to 2D deflagration scenario while both ODDT and CDDT models refer to 2D delayed-detonation scenario \citep{mae10}.}
\end{table}

Since measurement of the elemental abundances relative to hydrogen are limited to various uncertainties in the RGS spectral analysis (Appendix~\ref{sct:sys_err}), the number of stars ($N_{\rm ICM}^{li/m/d}$ in Equation~\ref{eq:z_icm}) in different enrichment channels (AGBs, SNcc and SNIa) are not well constrained. Thus, we turn to the abundance ratios (relative to Fe), which can be better constrained. The abundance ratios in the ICM can be characterized by
\begin{equation}
\frac{z_{i,~\rm ICM}}{z_{k,~\rm ICM}}=\frac{r_{\rm ICM}^{\rm li} <y_i^{\rm li}> +  <y_i^{\rm m}> + r_{\rm ICM}^{\rm d}y_i^{\rm d}}{r_{\rm ICM}^{\rm li} <y_k^{\rm li}> +  <y_k^{\rm m}> + r_{\rm ICM}^{\rm d}y_k^{\rm d}}~\frac{A_k~n_{k,~\rm ICM}}{ A_i~n_{i,~\rm ICM}}~,
\label{eq:ar_icm}
\end{equation}
where $k$ is the reference atom number (specifically refers to Fe $Z=26$ hereafter) and $r_{\rm ICM}^{\rm li/d} = N_{\rm ICM}^{\rm li/d} / N_{\rm ICM}^{\rm m}$.

\subsection{Origin of nitrogen enrichment}
\label{sct:origin}
\begin{figure}
\centering
\includegraphics[width=\hsize, trim={0.2cm 0.2cm 0.2cm 0.2cm}, clip]{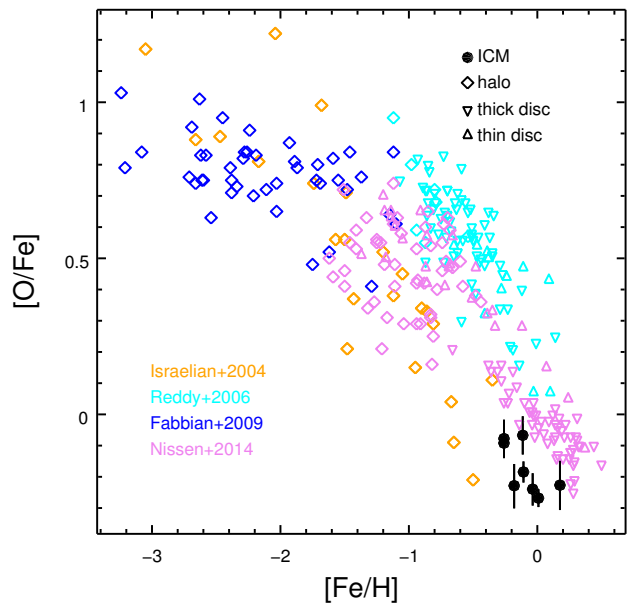}
\caption{The [O/Fe] -- [Fe/H] relation for ICM and the Galaxy. The ICM Fe abundances and O/Fe abundance ratios (Table~\ref{tbl:icm_all}) are shown as black dots with (statistical) error bars.The Galactic Fe abundances and O/Fe abundance ratios are taken from \citet{isr04} (halo, orange), \citet{red06} (disc and halo, cyan), \citet{fab09} (halo, blue) and \citet{nis14} (disc and halo, pink).}
\label{fig:sca_0826_2601}
\end{figure}

\begin{figure}
\centering
\includegraphics[width=\hsize, trim={0.2cm 0.2cm 0.2cm 0.2cm}, clip]{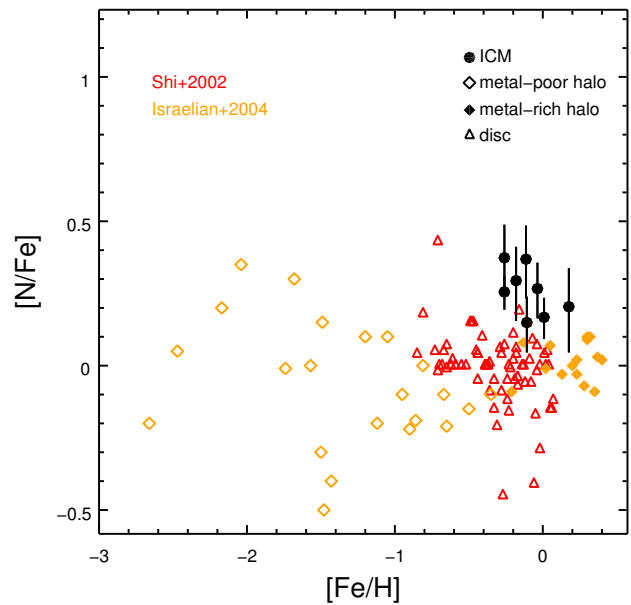}
\caption{Similar to Figure~\ref{fig:sca_0826_2601} but for the [N/Fe] -- [Fe/H] relation. The Galactic Fe abundances and N/Fe abundance ratios are taken from \citet{shi02} (disc, red triangles) and \citet{isr04} (halo, orange diamonds).}
\label{fig:sca_0726_2601}
\end{figure}

We first compare the abundance relations between the ICM and the Galaxy. Figure~\ref{fig:sca_0826_2601} and Figure~\ref{fig:sca_0726_2601} show the [O/Fe] -- [Fe/H] and [N/Fe] -- [Fe/H] relations, respectively. The corrections for non-local thermodynamic equilibrium (NLTE) and three dimensional (3D) stellar atmosphere models are not taken into account for some N and O abundances in the metal-poor halo stars ([Fe/H]$\lesssim -1$) in \citet{isr04} and \citet{spi05}. Detailed NLTE and 3D corrections \citep[see e.g.][for a review]{asp05}, are beyond the scope of this paper and do not alter our interpretation below.


For the [O/Fe] -- [Fe/H] relation (Figure~\ref{fig:sca_0826_2601}), a gradual decrease of [O/Fe] with increasing [Fe/H] is found in the [Fe/H]$\lesssim -1$ regime. 
This is due to the fact that the more metal-poor the SNcc progenitor is, the larger the [O/Fe] ratio in the SNcc yields \citep{rom10}. The rapid decrease of [O/Fe] in the [Fe/H]$\gtrsim-1$ regime, on the other hand, stems from the Fe enrichment by SNIa. The [O/Fe] ratio of the ICM is slightly smaller compared to disc stars 
in the Galaxy with the same [Fe/H] ratio. The overall [O/Fe]--[Fe/H] relation of the ICM and the Galaxy still supports the idea that they share the same enrichment channel (SNcc plus SNIa) for O and Fe. 

In contrast to the decreasing trend of [O/Fe] with increasing [Fe/H], a relatively flat [N/Fe] ratio with increasing [Fe/H] is found in Figure~\ref{fig:sca_0726_2601}, which indicates that N and O are enriched via different channels. In fact, the [N/Fe] -- [Fe/H] relation for the disk and halo stars can be explained \citep[see Fig.3 in][]{rom10} with AGB yields from \citet{kar10}. The [N/Fe] ratio of the ICM is slightly larger compared to halo stars in the Galaxy with the same [Fe/H] ratio. The overall [N/Fe]--[Fe/H] relation of the ICM and the Galaxy indicates that they share the same enrichment channel (AGB) for N.

\begin{figure}
\centering
\includegraphics[width=\hsize, trim={0.2cm 0.2cm 0.2cm 0.2cm}, clip]{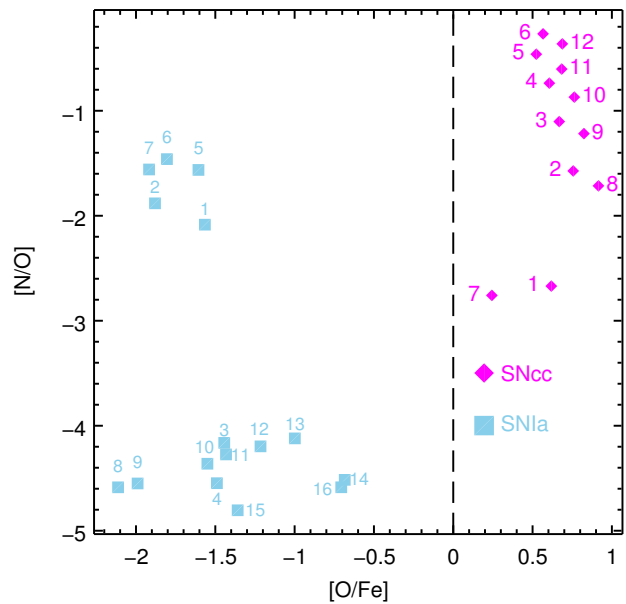}
\caption{The diamonds (magenta) are the IMF-weighted yields of SNcc (and PISNe for $Z_{\rm init}=0$), while the squares (blue) are SNIa yields. The indices next to the symbols indicate corresponding model dependency (Tabel~\ref{tbl:lim_idx} and Tabel~\ref{tbl:snia_idx}). The yields of all the elements from C to Zn can be found in Figure~\ref{fig:cf_wy_asncc_26} and \ref{fig:cf_snia_26}.
}
\label{fig:sne_070826}
\end{figure}

\begin{figure}
\centering
\includegraphics[width=\hsize, trim={0.2cm 0.2cm 0.2cm 0.2cm}, clip]{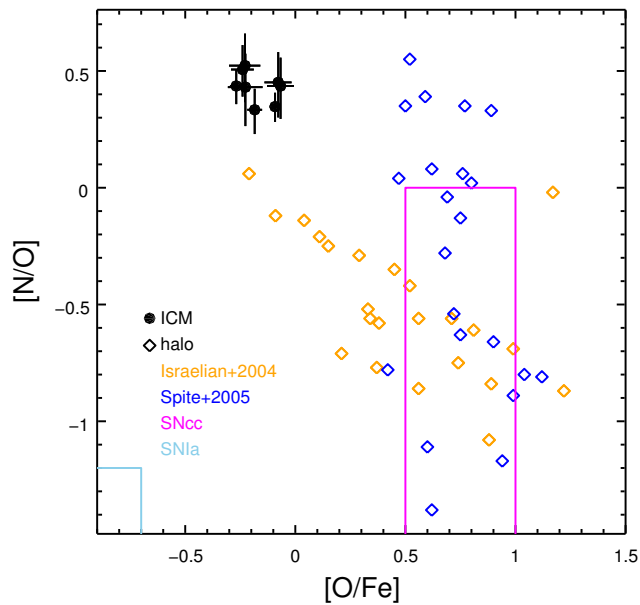}
\caption{Similar to Figure~\ref{fig:sca_0826_2601} but for [N/O] vs. [O/Fe].The results for Galactic stellar populations are taken from \citet{isr04, spi05} (halo, diamonds). The magenta box indicates the region of SNcc yields except for $Z_{\rm init} = 0$, while the blue box indicates the region of SNIa yields (Figure~\ref{fig:sne_070826}).}
\label{fig:sca_0708_0826}
\end{figure}

Secondly, we compare the [N/O]--[O/Fe] relation of supernova yields (Figure~\ref{fig:sne_070826}) to the observed abundances (Figure~\ref{fig:sca_0708_0826}).
The [O/Fe] ratio of the ICM is smaller compared to that of halo stars since the ICM is enriched by both SNcc and SNIa, while halo stars are mainly enriched by SNcc. Generally speaking, the [N/O] ratio in the ICM is larger than that of halo stars. Similar results have been reported in \citet{wer06a} for M\,87.  

If the chemical enrichment were completely due to massive stars ($N_{\rm ICM}^{\rm li/d} = 0$ in Equation~\ref{eq:z_icm}), then we would have [O/Fe]$\gtrsim0.5$ (Figure~\ref{fig:sne_070826}), except for $Z_{\rm init}=0$. For $Z_{\rm init}=0$, the [O/Fe] ratio can be lower than $\sim$0.5, due to the explosive O-burning by PISNe \citep{nom13}. In Figure~\ref{fig:sne_070826}, we assume all the very massive stars undergo PISNe (Section~\ref{sct:icm_ce}). In reality, 
the exact value of [O/Fe] (for $Z_{\rm init}=0$) might differ, depending on the IMF and the fraction of very massive stars that undergo PISNe. The [O/Fe] ratios in the ICM (Figure~\ref{fig:sca_0708_0826}) are in the range of (-0.5,~0.2), suggesting that the enrichment from SNIa is required for the ICM, unless PISNe contributes significantly.


The nitrogen enrichment via SNIa is negligible ([N/O]$\lesssim-1$). Therefore, one would expect [N/O] $\lesssim-0.2$ (Figure~\ref{fig:sne_070826}), if the chemical enrichment were completely due to massive stars ($N_{\rm ICM}^{\rm li/d} = 0$ in Equation~\ref{eq:z_icm}). We caution that the [N/O] ratio for $Z_{\rm init}=0$ in Figure~\ref{fig:sne_070826} is in fact a lower limit, since we do not include enrichment from metal-poor rotating massive stars before they explode as supernovae which is due to the lack of knowledge of corresponding number fraction and yields. \citet{chi06} have shown that a contribution, as large as [N/Fe]$\sim$0.5, from metal-poor ([Fe/H] $\lesssim$-2.5) rotating massive stars is required to solve the primary nitrogen problem \citep[see also Fig.3 in ][]{rom10}.
For a Salpeter IMF, the upper limit of [N/O] is estimated to be zero, given that not all the metal-poor massive stars are rotating (thus [N/Fe] $\lesssim0.5$)
and [O/Fe]$\gtrsim 0.5$ (Figure~\ref{fig:sne_070826}), regardless of $Z_{\rm init}$. The same upper limit of [N/O] holds for a top-heavy IMF with $Z_{\rm init} \gtrsim 0.001$. Nonetheless, for a top-heavy IMF with $Z_{\rm init} \lesssim 0.001$, the upper limit of [N/O] might be above zero, since [O/Fe] ratio can be lower than 0.5 as previously discussed.

The [N/O] ratios in the ICM are above zero at the $\gtrsim2.5\sigma$ confidence level (Table~\ref{tbl:icm_all}),  indicating that under a Salpeter IMF, the massive stars cannot be the main nitrogen factory. In this case, nitrogen mainly originates from low- and intermediate-mass stars (AGBs). Nevertheless, we cannot rule out that under a top-heavy IMF with a low initial metallicity ($Z_{\rm init}\lesssim0.001$), massive stars could be an important nitrogen enrichment factory.

Last but not least, we caution that the measured [N/Fe] and [N/O] ratios in Table~\ref{tbl:icm_all} might be biased. Due to the limited field of view (FOV) of RGS,  the abundance ratios obtained in the rather small ($\lesssim 0.05 r_{500}$) extraction regions do not necessarily represent the abundance patterns within the ``closed-box" (Section~\ref{sct:intro}). If the elements enriched via different channels were distributed into the ICM in different ways, so that, for instance, N were more centrally peaked than Fe and O, the resulting [N/Fe] and [N/O] ratios in the core region would appear to be larger. 

\subsection{Odd-$Z$ elements}
\label{sct:odd_z}
Previous studies on chemical enrichment in the ICM mainly focused on determining the SNIa fraction with respect to the total number of SNe that enriched the ICM \citep[e.g.][]{dpl06}. In terms of elemental abundances, most abundant even-$Z$ elements from oxygen up to and including nickel (except Ti) have been measured. Additionally, one odd-$Z$ Fe-peak element, Mn, is also studied in the stacked spectra of the CHEERS sample \citep{mer16a}. In terms of yields table, in addition to SNcc and SNIa, Pop III stars \citep{wer06b, dpl06} and Ca-rich gap transients \citep[CaRGT, ][]{mul14} have also been taken into account to interpret the observed abundance pattern. In this section, we include the nitrogen abundance and yields from AGBs \citep{cam08, kar10, nom13} for the chemical enrichment study of the ICM. 

Since the number of abundance ratios derived from the RGS spectra are rather limited, due to the relatively small coverage of the energy range, it is more meaningful when the abundance ratios measured with EPIC are also taken into account. Ideally, one needs to obtain the abundances within $\sim r_{500}$ of the ICM so that the ``closed-box" assumption is valid for massive clusters (Section~\ref{sct:intro}). In practice, especially for groups of galaxies, 
the FOV of RGS covers merely a tiny fraction of $r_{500}$. Moreover, the unknown nitrogen abundance gradients within $r_{500}$, prevent us from extrapolating the abundances out to $r_{500}$ with the obtained RGS abundances by hand.

We use both the RGS and EPIC results of NGC\,5044 (Table~\ref{tbl:ap_ngc5044}) for the exercise here, given that the measurement uncertainty of the nitrogen abundance is typical (neither too large nor too small), and the extraction regions are comparable  ($\sim$0.034~$r_{500}$ for RGS and $\sim$0.05~$r_{500}$ for EPIC). In Table~\ref{tbl:ap_ngc5044}, the N/Fe, O/Fe, Ne/Fe, Mg/Fe, and Ni/Fe abundance ratios are measured with RGS, while the Si/Fe, S/Fe, Ar/Fe, and Ca/Fe ratios are measured with EPIC (see details in Appendix~\ref{sct:epic}).

\begin{table}
\caption{The abundance ratios for NGC\,5044 within the extraction region (i.e. $\lesssim r/r_{500}$). Abundance ratios measured with EPIC spectra are labeled with $\dagger$. }
\label{tbl:ap_ngc5044}
\centering
\begin{tabular}{llcc}
\hline\hline
X/Fe & Value \\
\hline\hline
\noalign{\smallskip}
N/Fe & $1.4\pm0.3$ \\
O/Fe & $0.65\pm0.05$ \\
Ne/Fe & $0.68\pm0.08$ \\
Mg/Fe & $0.77\pm0.08$ \\
Si/Fe$^\dagger$ & $0.79\pm0.10$ \\
S/Fe$^\dagger$ &  $1.1\pm0.2$  \\
Ar/Fe$^\dagger$ & $1.0\pm0.3$  \\
Ca/Fe$^\dagger$ & $1.2\pm0.2$  \\
Fe & $0.72\pm0.02$   \\ 
Ni/Fe & $1.5\pm0.3$ \\ 
\hline
\end{tabular}
\end{table}

We emphasize that we focus on the comparison among different settings of the ICM enrichment model, i.e. the choice of IMF index and the initial metallicity of the stellar population, the choice of SNIa model, and whether to include enrichment from AGBs or not.

It is possible that SNIa from different channels (via single- or double-degenerate, deflagration or detonation, and super- or sub-Chandrasekhar limit) all play a role 
in the chemical enrichment of the ICM \citep{fin02, mer16b}. We only fit the measured abundance ratios with one set of SNIa yields for simplicity. Including an additional set of Ca-rich gap transients yields improves the statistics negligibly and does not change the above main points. 

\begin{figure}
\centering
\includegraphics[width=\hsize, trim={0.2cm 0.2cm 0.2cm 0.2cm}, clip]{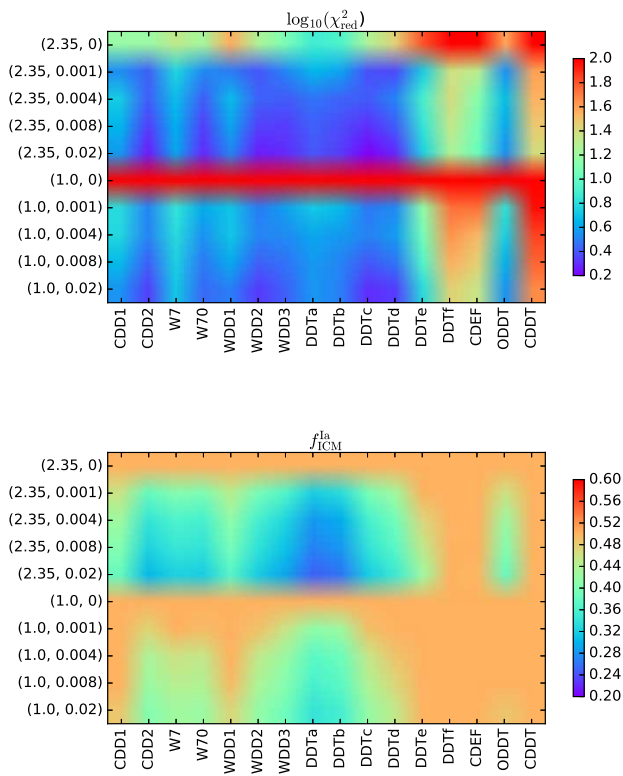}
\caption{Color map of reduced chi-squared ($\chi^2 / {\rm d.o.f.}$ in $\log_{10}$-scale, upper panel) and SNIa fraction ($f^{\rm Ia}_{\rm ICM}$, lower panel) for $10\times16$ combinations of yields we considered to fit the abundance ratios in NGC5044, without N/Fe (d.o.f. = 7). The X-axis labels indicate the SNIa models.
The Y-axis labels indicate IMF power-law index and the initial metallicity of the stellar populations.}
\label{fig:ngc5044_011_26_cm}
\end{figure}

Since almost all the measured abundance ratios (Table~\ref{tbl:ap_ngc5044}) are close to solar, we restrict the initial metallicity of stellar populations to be solar and sub-solar (i.e. excluding $Z_{\rm init}=0.05$). Thus, the observed abundance ratios are fitted to 10 (2 sets of IMF and 5 sets of $Z_{\rm init}$) $\times$16 (for SNIa) combinations of yield tables. The reduced chi-squared ($\chi_{\rm red}^2$) for all the fits are shown in the upper panel of Figure~\ref{fig:ngc5044_011_26_cm}. The $10\times16$ combinations of the chemical enrichment models are highly degenerate. We can reject a large number of combinations based on the statistics, say $\chi_{\rm red}^2 \gtrsim 3$, i.e. $\log_{10} (\chi_{\rm red}^2) \gtrsim 0.5$, however, the IMF power-law index and SNIa models cannot be exclusively obtained with current abundance measurements.  

\begin{figure}
\centering
\includegraphics[width=\hsize, trim={0.2cm 0.2cm 0.2cm 0.2cm}, clip]{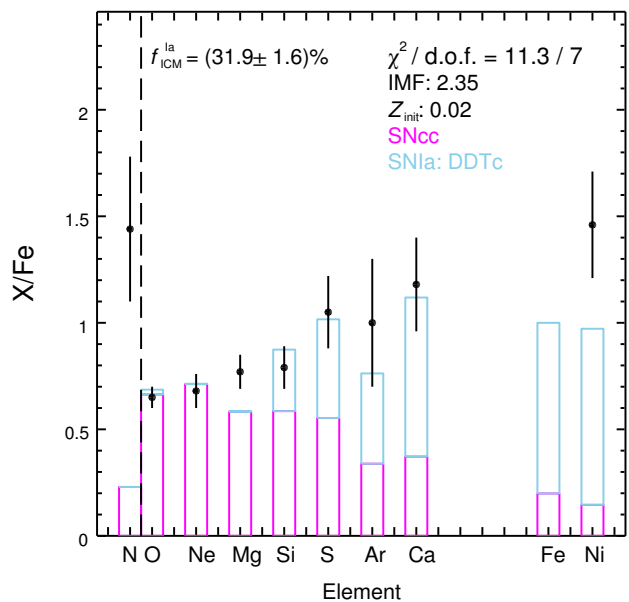}
\caption{One of the acceptable fits of chemical enrichment in NGC\,5044. Yields of SNcc (magenta) and SNIa (blue) are used for the fit with Equation~(\ref{eq:ar_icm}). The adopted IMF power-law index is 2.35 (Salpeter) and the initial metallicity of the stellar population is 0.02 (solar). N/Fe and yields of AGBs are not included in the fit, but shown for comparison. The SNIa fraction $f_{\rm ICM}^{\rm Ia} = r^{\rm d} / (r^{\rm d}+1)$ is $(31.9\pm1.6)\%$. }
\label{fig:ngc5044_011_26}
\end{figure}

\begin{figure}
\centering
\includegraphics[width=\hsize, trim={0.2cm 0.2cm 0.2cm 0.2cm}, clip]{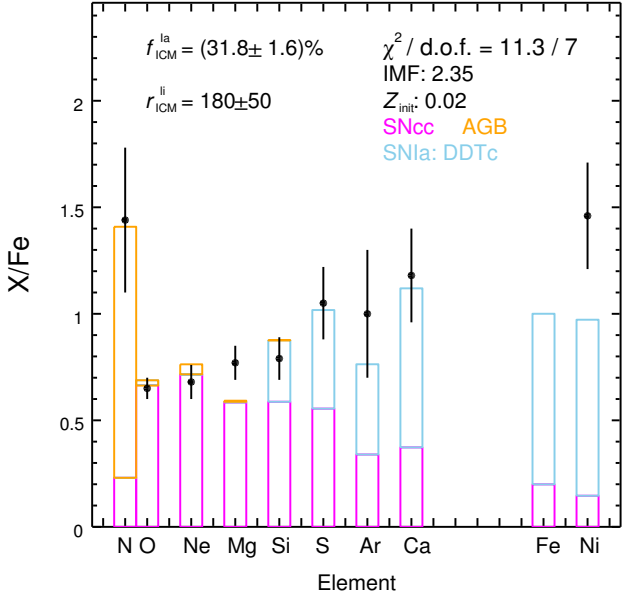}
\caption{Similar to Figure~\ref{fig:ngc5044_011_26} but N/Fe and AGB yields (orange) are included during the fit. Again, the favored IMF power-law index is 2.35 (Salpeter) and $Z_{\rm init}=0.02$. The SNIa fraction is consistent with the previous result. The ratio ($r_{\rm ICM}^{\rm li} = r^{\rm li} / r^{\rm m}$) between the number of low- and intermediate-mass stars and that of massive stars is $180\pm50$.}
\label{fig:ngc5044_111_26}
\end{figure}

Typical ``best" fits of the abundance ratios in NGC\,5044 to stellar yields are shown in Figure~\ref{fig:ngc5044_011_26} (without the N/Fe ratio and yields from AGBs) and Figure~\ref{fig:ngc5044_111_26} (with the N/Fe ratio and yields from AGBs). Without yields from AGBs (Figure~\ref{fig:ngc5044_011_26}), we also show the model prediction on the N/Fe ratio ($\sim0.2$). Compared to the measured N/Fe ratio ($1.4\pm0.3$), the predicted N/Fe ratio is lower by $\sim4\sigma$, indicating that the contribution from SNcc is not enough to explain the observed N/Fe ratio. When we include yields from AGBs (Figure~\ref{fig:ngc5044_111_26}), 
the predicted N/Fe ratio is consistent ($\lesssim 1\sigma$) with the observed N/Fe ratio. Additionally, the predicted O/Fe ratio decreases from $\sim$0.69 (SNe) to $\sim$0.66 (SNe + AGBs) due to the negative oxygen yields in AGBs.


In most cases, a Salpeter IMF provides better $\chi^2$ statistics (Figure~\ref{fig:ngc5044_011_26_cm}). When a Salpeter IMF and DDTc SNIa model are adopted, the SNIa fraction ($f_{\rm ICM}^{\rm Ia}$) is consistent with $\sim32\%$, whether we include N/Fe and AGBs enrichment or not (Figure~\ref{fig:ngc5044_011_26} and \ref{fig:ngc5044_111_26}). The ratio ($r_{\rm ICM}^{\rm li} = r^{\rm li} / r^{\rm m}$) between the number of low- and intermediate-mass stars and that of massive stars is $180\pm50$. Under a Salpeter IMF, the ratio ($r_{\rm ICM}^{\rm li}$) is expected to be $\sim40$, which is lower than the fitted value by $\sim3\sigma$. Again, the ``closed-box" assumption is not fulfilled here, so that if the AGB products were more centrally peaked than the SNcc products (Section~\ref{sct:origin}), a higher $r_{\rm ICM}^{\rm li}$ obtained here could be explained.

\begin{figure}
\centering
\includegraphics[width=\hsize, trim={0.2cm 0.2cm 0.2cm 0.2cm}, clip]{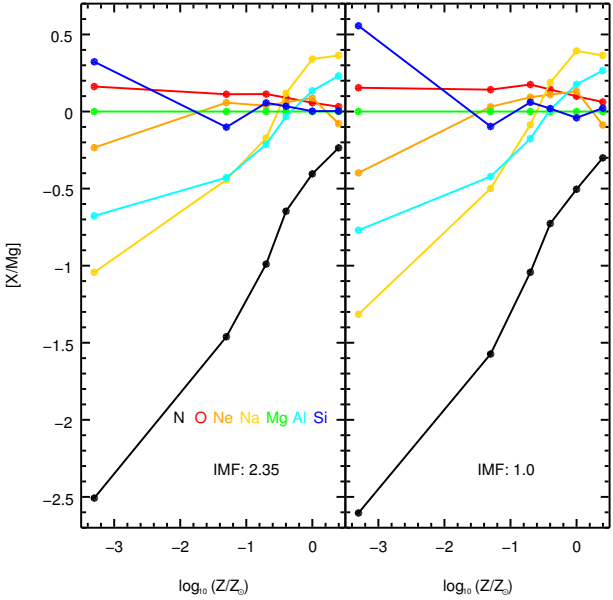}
\caption{The IMF weighted abundance ratios (with respect to Mg) as a function of the initial metallicity of massive stars (the SNcc channel). The results for $Z_{\rm init}=0$ are plotted at $\sim-3.3$. }
\label{fig:cf_wy_asncc_12}
\end{figure}

When N/Fe and AGB enrichment are not included in the fit, the ``best" fit initial metallicity is 0.02 (solar). This is mainly constrained by the less than unity O/Mg abundance ratio (Figure~\ref{fig:cf_wy_asncc_12}). Including N/Fe and AGB enrichment again favours solar initial metallicity. We also notice that, in both cases, a wide range of $Z_{\rm init}$ yields comparable results (Table~\ref{tbl:ngc5044_26_af}), except for $Z_{\rm init} = 0$.

In principle, when odd-$Z$ elemental abundances, like nitrogen, are included in the analysis, the initial metallicity of the stellar population should be better constrained, since yields of odd-$Z$ elements increase significantly with increasing $Z_{\rm init}$ owing to a surplus of neutrons \citep{nom13}, while those of even-$Z$ and Fe-peak elements are almost constant over a wide range of metallicities. This is shown clearly in Figure~\ref{fig:cf_wy_asncc_12} for massive stars. 
We emphasize that the denominator of the abundance ratio on the Y-axis is set to Mg instead of Fe in Figure~\ref{fig:cf_wy_asncc_12}. This is because Mg enrichment via SNIa and AGBs are negligible, so that the observed abundance ratios of Na/Mg and Al/Mg can be used directly to probe the initial metallicity of the stellar population.  

\begin{table}
\caption{The ``best" fits (d.o.f. = 7) of chemical enrichment in NGC\,5044, given the IMF power-law index and initial metallicity of the stellar population.}
\label{tbl:ngc5044_26_af}
\centering
\begin{tabular}{cccccccccccccccccccccc}
\hline\hline
IMF$^{a}$ & $Z_{\rm init}^{b}$ & SNIa  & AGB$^{c}$ &  $\chi^2$ \\
\hline\hline
\noalign{\smallskip}
2.35 & 0.02 & DDTc & N & 11.3  \\
2.35 & 0.008 & DDTc & N & 14.5  \\
2.35 & 0.004 & WDD3 & N & 18.1  \\
2.35 & 0.001 & DDTd & N & 16.5  \\
2.35 & 0 & DDTa & N & 53.9  \\
\noalign{\smallskip}
1.0 & 0.02 & DDTc & N & 14.2  \\
1.0 & 0.008 & DDTc & N & 18.0  \\
1.0 & 0.004 & DDTc & N & 22.7  \\
1.0 & 0.001 & DDTc & N & 21.5  \\
1.0 & 0 & DDTa & N & >100  \\
\noalign{\smallskip}
2.35 & 0.02 & DDTc & Y & 11.3  \\
2.35 & 0.008 & DDTc & Y & 13.2  \\
2.35 & 0.004 & DDTc & Y & 14.5  \\
2.35 & 0.001 & DDTd & Y & 22.5 \\
2.35 & 0 & DDTa & Y & 52.5  \\
\noalign{\smallskip}
1.0 & 0.02 & DDTc & Y & 13.7  \\
1.0 & 0.008 & DDTc & Y & 15.7  \\
1.0 & 0.004 & DDTc & Y & 18.7  \\
1.0 & 0.001 & DDTc & Y & 18.4 \\
1.0 & 0 & DDTa & Y & >100  \\
\hline
\end{tabular}
\tablefoot{$^{a}$ The power-law index of the IMF. $^{b}$ The initial metallicity of the stellar populations. $^{c}$ Whether the N/Fe ratio and the yields of AGBs 
are included in the fit.}
\end{table}

For future work, more accurate abundance measurements of odd-$Z$ elements including N, Na and/or Al are required to better constrain the initial metallicity of the stellar population. Current instruments lack the spectral resolution to resolve the Ly$\alpha$ lines of \ion{Na}{xi} and \ion{Al}{xiii}. Hopefully, future missions with high spectral resolution and large effective area like {\it XARM} (X-ray astronomy recovery mission) and {\it Athena} \citep{nan13} will address this issues.

\section{Conclusions}
We constrain the N/Fe ratio in the core ($r/r_{500}\lesssim0.5$) of one cluster (A\,3526) and seven groups of galaxies (M\,49, M\,87, NGC\,4636, NGC\,4649, NGC\,5044, NGC\,5813, NGC\,5846) in the CHEERS sample with high-resolution RGS spectra. Our main conclusions are summarized as follows: 
\begin{enumerate}
\item The nitrogen abundance is well constrained ($\gtrsim3\sigma$) in objects with a relatively cool ICM ($kT\lesssim 2$~keV). For some of the systems (e.g. NGC\,3411) in the CHEERS sample, more exposure time is required to better constrain the nitrogen abundance. In objects with a hotter ICM ($kT\gtrsim2-3$~keV), the continuum level is high so that weak emission lines like \ion{N}{vii} Ly$\alpha$ cannot be well constrained.  

\item Both the [O/Fe]--[Fe/H] and [N/Fe]--[Fe/H] relations observed in the ICM are comparable to those observed in different stellar populations in the Galaxy,
indicating that the enrichment channels for N, O and Fe are expected to be the same. One possible explanation for the super solar N/Fe and N/O ratios in the ICM
is the bias introduced by our small extraction region ($r < 0.05 r_{500}$). This potential bias can be confirmed by radial abundance maps for N, O, and Fe in future work. 

\item If the observed ratio [N/O]$>0$ (at the $\gtrsim2.5\sigma$ confidence level) is not biased due to the small extraction region, under a Salpeter IMF, the low- and intermediate-mass stars are found to be the main metal factory for nitrogen. This is in agreement with the Galactic chemical evolution theory and previous studies of M 87. Nitrogen enrichment from massive stars might still be important, especially if the stellar population would have a top-heavy IMF and zero initial metallicity.  

\item We find the obtained SNIa fraction is insensitive to the N abundance and AGB yields. 

\item We also point out that accurate abundance measurements of odd-$Z$, such as N, Na, and Al can certainly help to better constrain the initial metallicity of the stellar population that enriched the ICM.

\end{enumerate}

\begin{acknowledgements}
This paper is dedicated to the memory of our deeply valued colleague Yu-Ying Zhang, who recently passed away. This work is based on the \textit{XMM-Newton} AO-12 proposal (ID: 72380) ``The \textit{XMM-Newton} view of chemical enrichment in bright galaxy clusters and groups" (PI: de Plaa). The observations are obtained with XMM-{\it Newton}, an ESA science mission with instruments and contributions directly funded by ESA member states and the USA (NASA).SRON is supported financially by NWO, the Netherlands Organization for Scientific Research. J.M. grateful acknowledges discussions and consultations with C. de Vries, J. Sanders and O. Pols. C.P. acknowledges support from ERC Advanced Grant Feedback 340442. Y.Y.Z. acknowledges support by the German BMWi through the Verbundforschung under grant 50OR1506. H.A. acknowledges the support of NWO via a Veni grant. 
\end{acknowledgements}

\begin{appendix}
\section{Global spectral fit}
\begin{figure*}
\centering
\includegraphics[width=\hsize, trim={1cm 2cm 1cm 2cm}, clip]{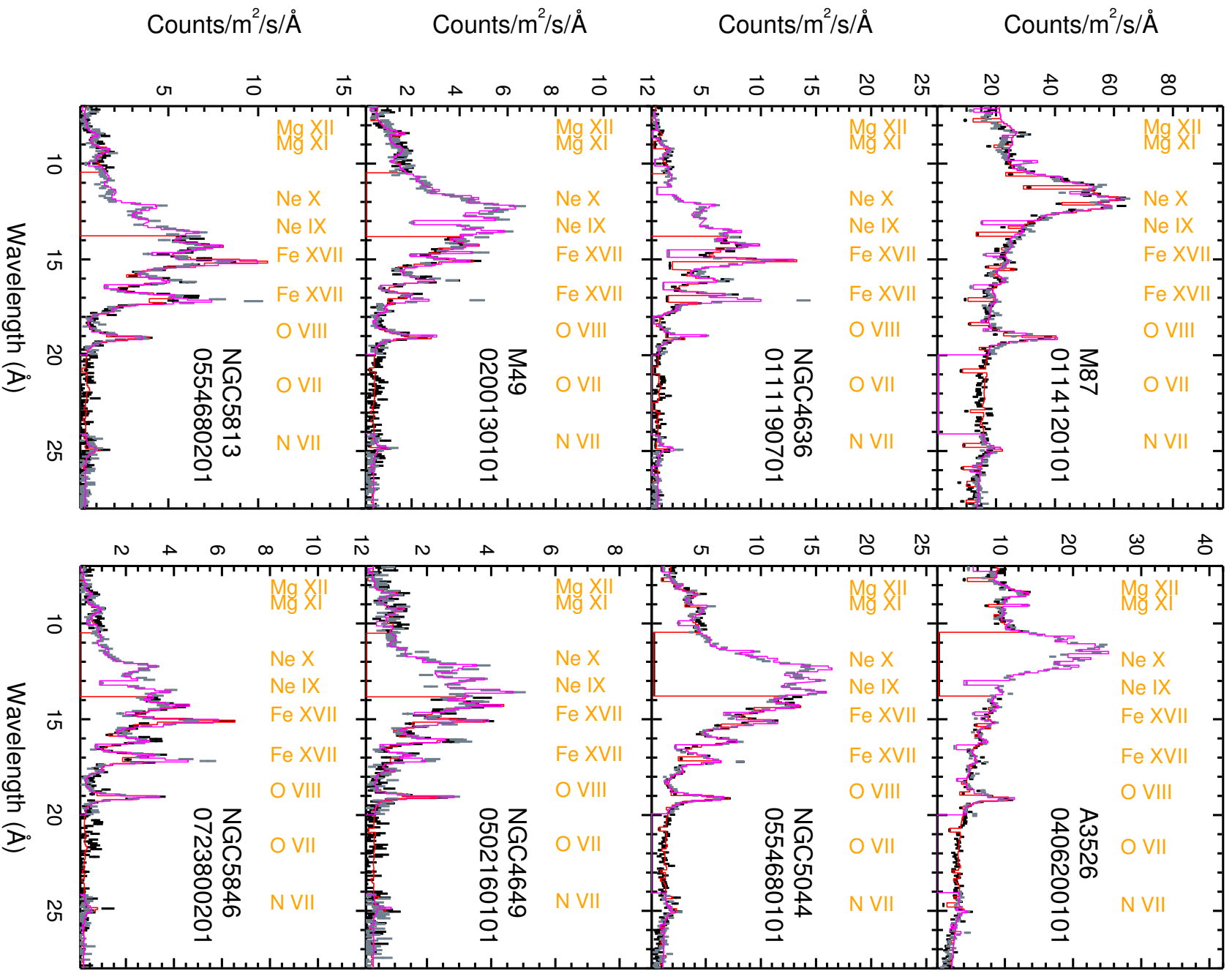}
\caption{The global fits to the 7--28 $\angstrom$ wavelength range. The data points are shown in black (RGS1) and grey (RGS2) and the model spectra are shown in red (RGS1) and magenta (RGS2) histograms. Spectra from merely one observation per target are shown for clarity. }
\label{fig:cheers_spec_plot}
\end{figure*}

The global fits to the 7--28 $\angstrom$ wavelength range for each source in Table~\ref{tbl:icm_all} are shown in Figure~\ref{fig:cheers_spec_plot}. The location (in the observed frame) of  characteristic emission lines are labeled, including the Ly$\alpha$ line from H-like \ion{N}{vii}, \ion{O}{viii}, \ion{Ne}{x}, \ion{Mg}{xii}, He-like triplets from \ion{O}{vii}, \ion{Ne}{ix}, \ion{Mg}{xi} and the resonance and forbidden lines of Ne-like \ion{Fe}{xvii}. 

\label{sct:global_plot}

\section{Systematic uncertainties in spectral analysis}
\label{sct:sys_err}
\subsection{Differential emission measure distribution}
\label{sct:dem}
Fitting a multi-temperature plasma with a single temperature (1T) model would often over-estimate the emission measure and under-estimate the abundances \citep{buo94, buo00}. Alternatively, a two-temperature (2T) or multi-temperature (GDEM) model can measure the nitrogen abundance more accurately.

\begin{figure}
\centering
\includegraphics[width=\hsize, trim={0.2cm 0.2cm 0.2cm 0.2cm}, clip]{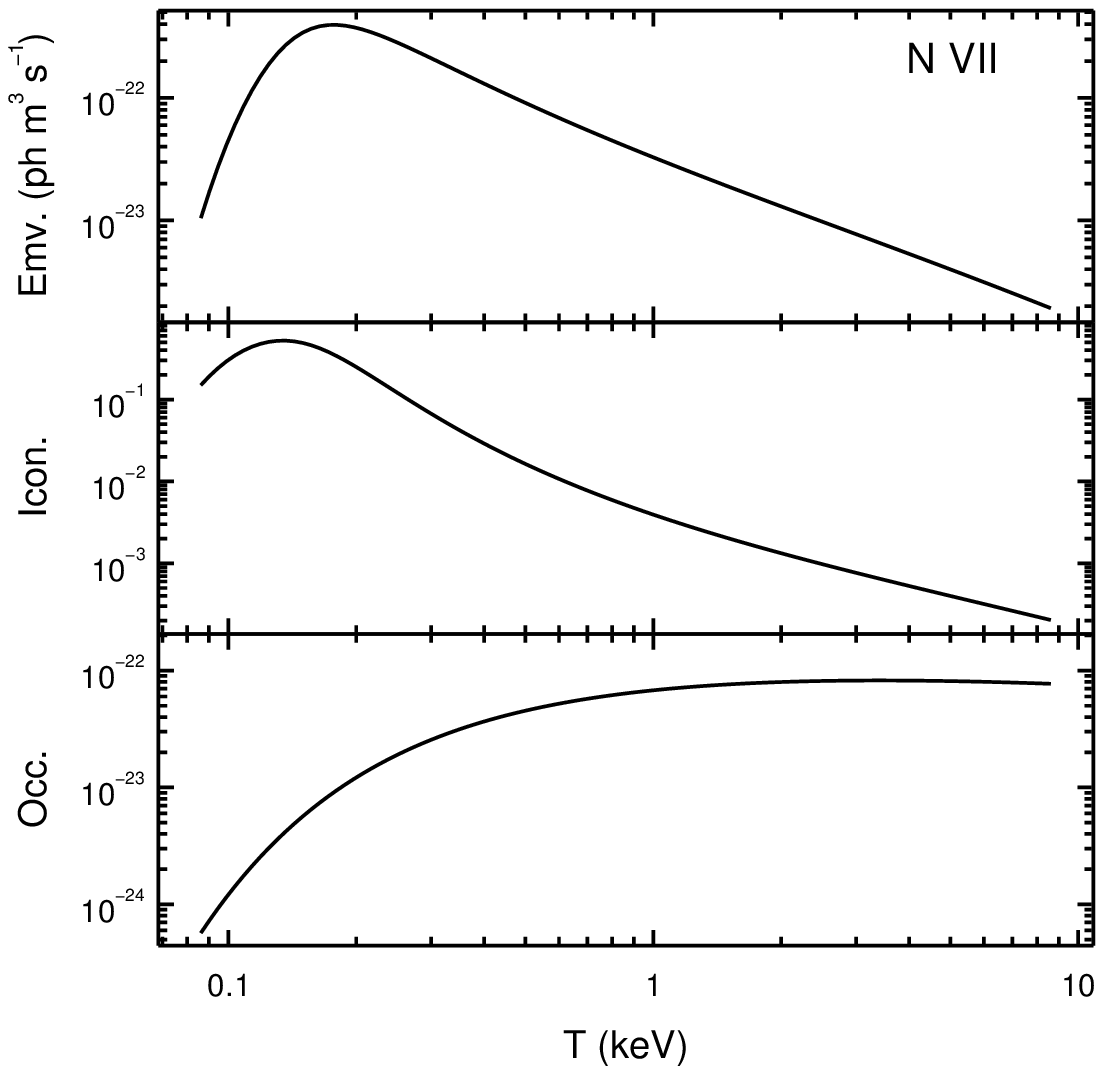}
\caption{The \ion{N}{vii} Ly$\alpha$ line emissivity (top), relative ion concentration (middle) and relative level occupation (bottom) of the two upper levels 
$^2 P_{0.5}$ and $^2 P_{1.5}$ (in sum, since the fine-structure lines cannot be distinguished). The underlying ionization balance is \citet{urd17} and the proto-solar abundance of \citet{lod09} is used. In a hot ($kT\gtrsim 0.6$~keV) single-temperature CIE plasma, nitrogen is almost fully ionized in the form of \ion{N}{viii}, i.e. the ion concentration of \ion{N}{viii} is $\sim 1$. Most of the \ion{N}{vii} is in the ground level $^1 S_{0.5}$, i.e. the level occupations of  $^1 S_{0.5}$ is close to unity.}
\label{fig:icon_occ}
\end{figure}

In a 2T model, if the emission measure of the hotter component is $\gtrsim5$ times that of the cooler one, the \ion{N}{vii} in the hotter CIE component contributes 
more to the observed \ion{N}{vii} Ly$\alpha$ emission in the spectra. Given the same emission measure and nitrogen abundance, the \ion{N}{vii} Ly$\alpha$ line flux is proportional to the \ion{N}{vii} ion concentration (relative to all the nitrogen atoms and ions in the ICM) times the level occupations of $^2 P_{0.5}$ and $^2 P_{1.5}$ (the sum of occupations of all the levels are defined as unity). As the level occupations increase gradually as a function of plasma temperature (bottom panel in Figure~\ref{fig:icon_occ}), the \ion{N}{vii} ion concentration is the leading factor to determine the line emissivity. As mentioned above, we tie the abundances in our 2T model, thus, assuming $kT_{\rm c} \lesssim 0.7$~keV and $kT_{\rm h} \gtrsim 2$~keV, when $Y_{\rm c} / Y_{\rm h} \lesssim 0.2$, 
the \ion{N}{vii} in the hotter component contributes more to the emission line, while for $Y_{\rm c} / Y_{\rm h} \gtrsim 0.2$, the \ion{N}{vii} is mainly from the cooler component.

In addition, the line emissivity of \ion{N}{vii} Ly$\alpha$ peaks around $T \sim 2\times 10^6$~K \citep{kaa08}, implying that nitrogen is preferably found in relatively cooler plasma. As the line emissivity declines rapidly with the increasing temperature of the plasma (top panel in Figure~\ref{fig:icon_occ}), we find it is rather difficult to well constrain the nitrogen abundance via the extremely weak \ion{N}{vii} Ly$\alpha$ emission line embedded in the relatively high continuum where $kT\gtrsim 2-3$~keV. 

\subsection{Spatial broadening model}
\label{sct:lpro}
The spatial broadening model {\it lpro} is built based on the spatial broadening profile. The latter is obtained from the MOS1 image, since the MOS1 DETY direction is in parallel to the RGS1 dispersion direction. There are two more free parameters in {\it lpro}, the scaling factor ($s$) and the offset parameter. Here we discuss the systematic uncertainties of the spatial broadening model. 

For instance, in M\,87, due to the presence of the bright non-thermal emission in the second observation (ObsID: 0200920101), not only the spectra are heavily contaminated, but also the spatial broadening model created with the MOS1 image is affected. The brighter the central non-thermal emission, the more centrally peaked the surface brightness profile (seen indirectly in Figure~\ref{fig:vprof_plot}). Spatial broadening models built on these biased surface brightness profiles 
reflect no longer the proper spatial extent of the ICM.

\begin{figure}
\centering
\includegraphics[width=\hsize, trim={0.2cm 0.2cm 0.2cm 0.2cm}, clip]{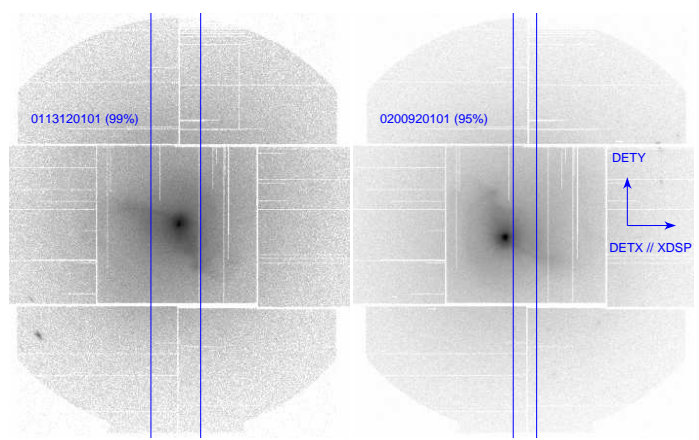}
\caption{The 7--28~$\angstrom$ energy band MOS1 image in the detector coordinate system for M87. The MOS1 DETX axis is in parallel with the cross dispersion direction (XDSP) of RGS. In the second observation (0200920101), the non-thermal emission is much brighter,  offset by $\sim$1.5 arcmin and the image is rotated $\sim$188 degrees clockwise. The (blue) rectangular boxes indicated the extraction regions for {\it rgsvprof}. For the first observation, a 99\%-xpsf ($\sim$3.4~arcmin) extraction region, aligned with the RGS source extraction region, is used. For the seconding observation, only the onset 95\%-xpsf ($\sim$1.6~arcmin) extraction region (to avoid central contamination) is shown for clarity.}
\label{fig:img_det_m87}
\end{figure}

\begin{figure}
\centering
\includegraphics[width=\hsize, trim={0.2cm 0.2cm 0.2cm 0.2cm}, clip]{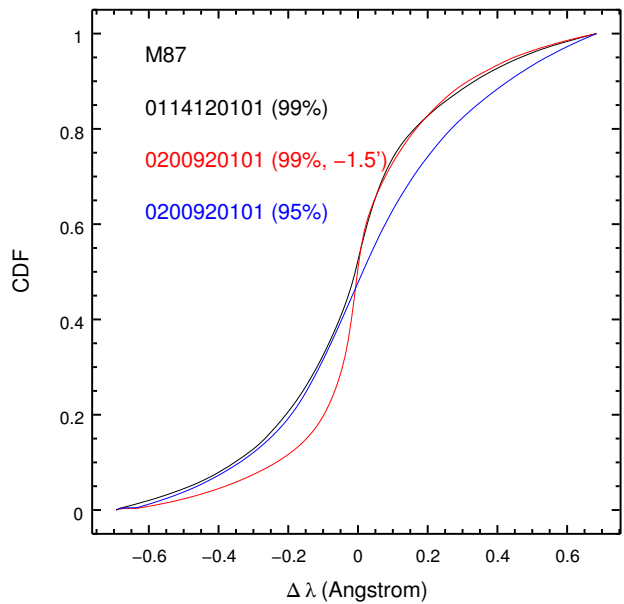}
\caption{The cumulative distribution function (CDF) of the spatial broadening profile of M87. We built the spatial broadening profiles for both two observations (black and red) of the RGS 99\%-xpsf source extraction regions. For the second observation (0200920101), a 1.5 arcmin offset is applied so that the extraction region 
is centred on the peak of the X-ray emission. Moreover, for the second observation, we also built a spatial broadening profile with a narrower 95\%-xpsf extraction region (blue) suffering less from central contamination (Figure~\ref{fig:img_det_m87}).}
\label{fig:vprof_plot}
\end{figure}

We compare the (global) fit results using different line broadening profile of M\,87 here. If the non-thermal emission were merely a point source and the ICM were azimuthally symmetric, one might fit the observed 2D image with two Gaussian/Lorentzian  profiles with different widths, then subtract the non-thermal emission counterpart to obtain the profile for the ICM only. However, this is not the case for M\,87 due to its azimuthal asymmetry (Figure~\ref{fig:img_det_m87}). 
Because the emission centre is offset by $\sim$1.5~arcmin in 0200920101, we took advantage of a $\sim$1.6-arcmin-wide extraction region without the non-thermal emission, leading to a better yet still biased (probably flatter) spatial broadening model (Figure~\ref{fig:vprof_plot}). Whereas, we found the {\it lpro} scaling factor (free parameter) can account for the bias in the spatial broadening profile.


Other than the accuracy of the spatial broadening profile, the scaling factors might be different for different thermal components and/or different ions within the same thermal component. When studying the O VII He-like triplets in the CHEERS sample, \citet{pin16} found the spatial extent of the cooler ICM component
is narrower than that of the hotter counterpart, by using two {\it lpro} model components for the two temperature components. Since in most cases, nitrogen from the hotter component contributes the most to the emission line we observed, thus, applying the same {\it lpro} model component (mainly determined by the high-temperature lines) to both the hotter and the cooler thermal component should be fine in our case.

\subsection{RGS background model}
\label{sct:mbkg}
In some cases in the CHEERS sample,  the modelled background level is even higher than the source continuum level at $\lambda \gtrsim 20~\angstrom$ (Figure~\ref{fig:cheers_spec_plot_zoom}). Thus we check the systematic uncertainties of the modelled background as well. We use A\,2029 as an example to compare the observed spectra from an offset observation toward A\,2029 with the RGS modelled background. 

The outskirts of A\,2029 were observed with XMM-{\it Newton} in 2015. The projected angular distances for the outskirts are $\sim 20$~arcmin, i.e. at least $\sim 1.3~r_{500}$. The outskirts of A\,2029 were also observed by {\it Suzaku} and no statistically significant emission is detected beyond 22~arcmin, except for the northern observation \citep{wal12}. That is to say, the spectra of the observations toward the outskirts of A\,2029 can be considered as background spectra. We used the same data reduction method described in Section~\ref{sct:dr} to screen out the flare time intervals and
extracted the RGS spectra in the 99\%-xpsf extraction region.

\begin{figure}
\centering
\includegraphics[width=\hsize, trim={0.2cm 0.2cm 0.2cm 0.2cm}, clip]{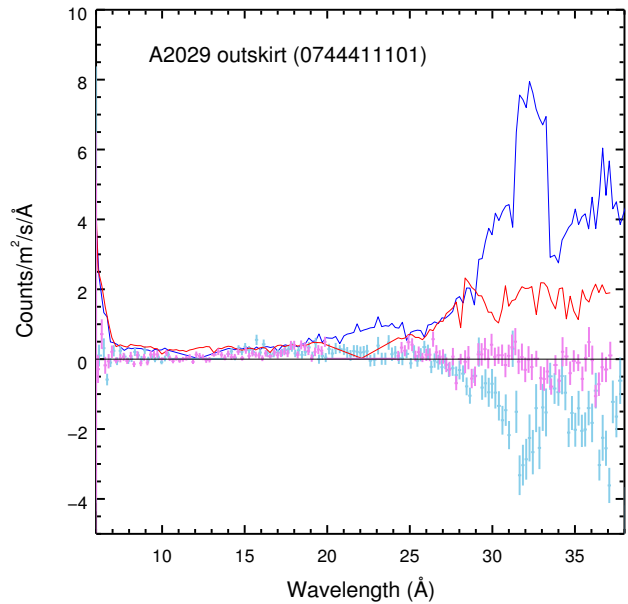}
\caption{The RGS 1st-order spectra of A\,2029 outskirt (ObsID: 0744411001).
The data with error bars (light blue for RGS1 and violet for RGS2) are the observed background spectra minus the modelled background spectra, which are expected to be around zero, if the modelled background spectra are accurate. The solid lines (deep blue for RGS1 and red for RGS2) are the (subtracted) modelled background spectra obtained with {\it rgsproc}.}
\label{fig:bkg_a2029}
\end{figure}

In Figure~\ref{fig:bkg_a2029}, we plot the RGS 1st-order ``net" (observed minus modelled background) spectra of the A\,2029 southern outskirt (ObsID: 0744411001). If the modelled background spectra is accurate enough, the ``net" spectra should be consistent with zero. Above $\sim$26.5~$\angstrom$, we see
the modelled background spectrum of RGS1 is significantly overestimated. The RGS2 modelled background is more accurate than that of RGS1
above $\sim$26.5~$\angstrom$. Therefore, for any source with redshift $z\gtrsim 0.07$, the accuracy of the RGS1 modelled background can be an issue for the \ion{N}{vii} Ly$\alpha$ line measurement, if the modelled background level dominates the source continuum level for (redshifted) $\lambda \gtrsim 26.5~\angstrom$.

Last but not least, we take NGC\,5846 as an example to show the impact on the abundance measurement if the modelled background were systematically over- or under-estimated. We used the FTOOL task {\it fcalc} to increase/decrease the values of the entire BACKSCAL column by 10\% for the modelled background spectra FITS file. Then we re-analyse the source spectra after subtracting the modified modelled background spectra. Given a 2T model, compared to the results of unmodified modelled background (Table~\ref{tbl:icm_all}), the Fe abundance increases/decreases dramatically by $+0.33$ (for 90\% BACKSCAL) and $-0.15$ (for 110\% BACKSCAL), respectively. The deviations are significantly larger compared to pure statistical errors. Nevertheless, the abundance ratios of N/Fe and O/Fe are consistent with $3.4$ and $1.3$, respectively. That is to say, the abundance ratios are robust given a 10\% (constant) uncertainty in the modelled background spectra. Similar checks are also performed on other sources with higher modelled background level. 

\section{EPIC spectral analysis of NGC\,5044 with SPEX v3.03}
\label{sct:epic}
The EPIC Si/Fe, S/Fe, Ar/Fe, and Ca/Fe abundance ratios ($\dagger$ in Table~\ref{tbl:ap_ngc5044}) have been reported in \citet[][their Table D.1]{mer16a}. However, an older version of SPEX (v2.05) was used at that time. In the present work, we reanalyze the EPIC spectra with SPEX v3.03. As shown in Table~\ref{tbl:ngc5044_epic}, the newly obtained abundance ratios are consistent (at a 1$\sigma$ confidence level) with those reported in \citet{mer16a}. 

\begin{table}
\caption{The best-fit results of the EPIC spectra of NGC\,5044 using SPEX v2.05 and v3.03. MOS and pn spectra are fitted simultaneously. }
\label{tbl:ngc5044_epic}
\centering
\begin{tabular}{ccccc}
\hline\hline
\noalign{\smallskip}
SPEX & v2.05 & v3.03 & v3.03 \\
\noalign{\smallskip}
SPEXACT & v2.05 & v2.05 & v3.03 \\
\hline
\noalign{\smallskip}
Model & GDEM & 3T & 3T \\
\noalign{\smallskip}
$C$-stat & 6502 & 6210 & 5635 \\
\noalign{\smallskip}
d.o.f. & 1512 & 1287 & 1287 \\
\noalign{\smallskip}
Norm. & $2.153\pm0.014$ & $2.218\pm0.008$ & $2.077\pm0.023$\\
\noalign{\smallskip}
$kT$ & $0.974\pm0.002$ & $1.043\pm0.003$ & $0.962\pm0.004$ \\
\noalign{\smallskip}
Si/Fe & $0.93\pm0.14$ & $0.96\pm0.07$ & $0.79\pm0.10$\\
\noalign{\smallskip}
S/Fe & $1.3\pm0.3$ & $1.4\pm0.1$ & $1.1\pm0.2$ \\
\noalign{\smallskip}
Ar/Fe & $1.4\pm0.5$ & $1.3\pm0.4$ & $1.0\pm0.3$ \\
\noalign{\smallskip}
Ca/Fe & $1.5\pm0.3$ & $1.6\pm0.3$ & $1.2\pm0.2$ \\
\hline
\end{tabular}
\tablefoot{The normalization in units of $10^{71}~{\rm m^{-3}}$ refers to the total emission measure. The temperature (in keV) here is where the differential emission measure reaches its maximum.}
\end{table}

More accurate and complete atomic data (SPEXACT v3.03) are used in SPEX v3.03 (Section~\ref{sct:spec_mo}). The total number of lines has increase by a factor of $\sim400$ to reach about 1.8 million in the new version. Consequently, multi-temperature plasma models like GDEM, which has about 20 different normalization/temperature components, is computational expensive for SPEX v3.03. A three temperature (3T) model would be a cheaper alternative for SPEX v3.03. To mimic a Gaussian differential emission measure distribution (GDEM) with a three temperature distribution, we set the temperatures of all three components to be free, the normalization of the main component is also allowed to vary, while the normalization of the low- and high-temperature components are fixed to be half of that of the main component \citep[see also][]{mer18}.

MOS and pn spectra are fitted simultaneously. When we use the old atomic database (SPEXACT v2.05), the best-fit $C$-stat to degrees of freedom ratios are 6502/1512 (GDEM in SPEX v2.05) and 6210/1287 (3T in SPEX v3.03), respectively. As expected, the ratio is slightly worse for the 3T model. The degrees of freedoms are different mainly due to the fact that the optimal binning algorithm \citep{kaa16} is different in the two versions of SPEX.

When we use the 3T model and SPEX v3.03, the best-fit $C$-stat to degrees of freedom ratios are 6210/1287 (SPEXACT v2.05) and 5635/1287 (SPEXACT v3.03), respectively. This shows the improvement with the new atomic data.

\section{IMF weighted SNcc yields and yields of SNIa.}
\label{sct:yields}
\begin{figure}
\centering
\includegraphics[width=\hsize, trim={0.2cm 0.2cm 0.2cm 0.2cm}, clip]{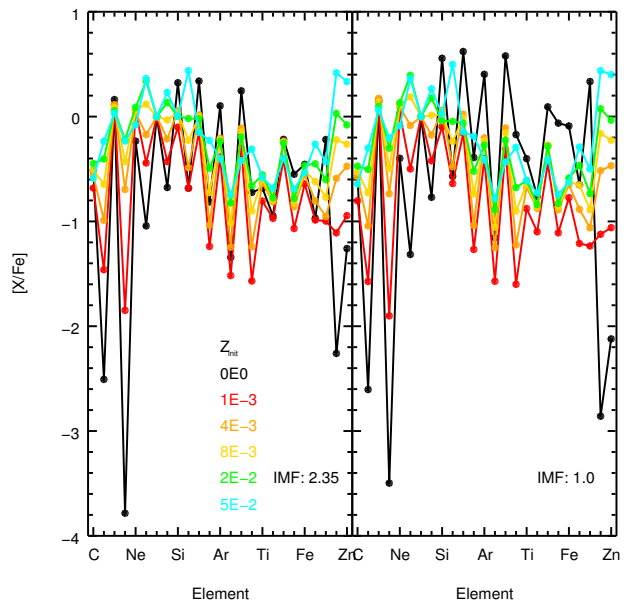}
\caption{IMF weighted SNcc yields, based on the yields table of \citet{nom13}. }
\label{fig:cf_wy_asncc_26}
\end{figure}

\begin{figure}
\centering
\includegraphics[width=\hsize, trim={0.2cm 0.2cm 0.2cm 0.2cm}, clip]{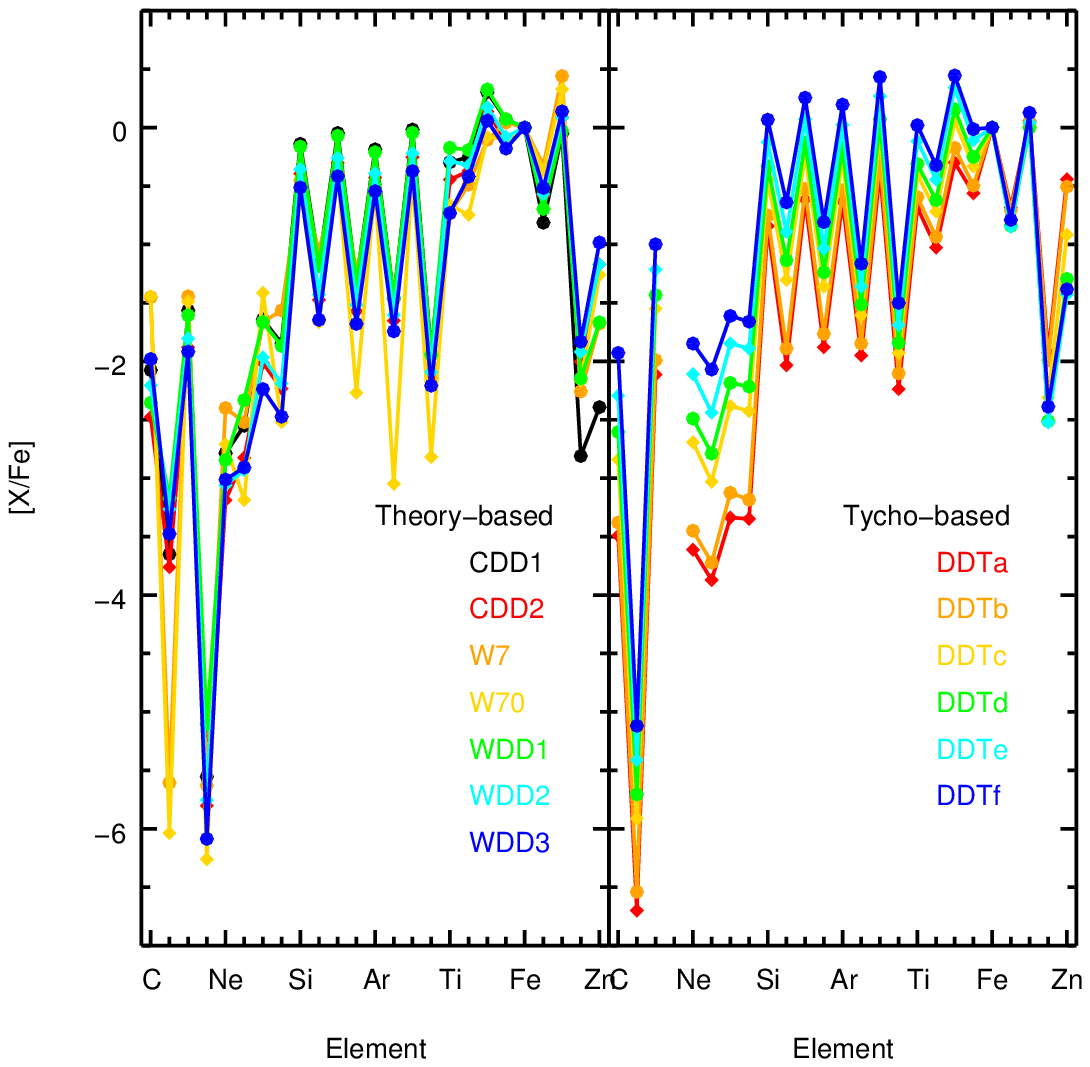}
\caption{Yields from various SNIa models \citep{iwa99,bade06}.}
\label{fig:cf_snia_26}
\end{figure}

IMF weighted core-collapse supernovae (SNcc) yields are shown in Figure~\ref{fig:cf_wy_asncc_26} with different initial metallicity ($Z_{\rm init}$) and initial mass function (IMF) for the stellar progenitors. Yields table of \citet{nom13} are used for the calculation. Figure~\ref{fig:cf_snia_26} shows the Type Ia supernovae (SNIa) yields based on theoretical calculations \citep{iwa99} and observations of the Tycho supernova remnant \citep{bade06}.

\end{appendix}

\end{document}